\documentclass[
amsmath,amssymb,prd, aps, a4paper, reprint, superscriptaddress,
nofootinbib,
floatfix
]{revtex4-2}
\usepackage[english]{babel}
\usepackage[T1]{fontenc}
\usepackage[utf8]{inputenc}
\usepackage[normalem]{ulem}

\usepackage{amsthm}
\usepackage{mathtools}
\usepackage{physics}
\usepackage{xcolor}
\usepackage{graphicx}
\usepackage[left=23mm,right=13mm,top=35mm,columnsep=15pt]{geometry} 
\usepackage{adjustbox}
\usepackage{placeins}
\usepackage{csquotes}
\usepackage{subcaption}

\usepackage{bm} 
\usepackage{balance}
\usepackage[pdftex, pdftitle={Article}, pdfauthor={Author}]{hyperref} 
\usepackage{stmaryrd}
\usepackage{tabulary}

\usepackage{txfonts}
\usepackage{orcidlink}

\usepackage{dcolumn}
\usepackage{bm}

\usepackage{comment}

\usepackage{cellspace} 
\newcommand{\observed}[1]{\tilde{#1}}
\newcommand{\BNT}[1]{\hat{#1}}

\begin{document}

\preprint{APS/123-QED}

\title{Lensing without mixing: Probing baryonic acoustic oscillations and other scale-dependent features in cosmic shear surveys}

\author{David Touzeau}%
\email{david.touzeau@ipht.fr}
\affiliation{Université Paris-Saclay, CNRS, CEA, Institut de physique théorique, 91191, Gif-sur-Yvette, France}
\author{Alexandre Barthelemy}
\affiliation{Université Paris-Saclay, Université Paris Cité, CEA, CNRS, AIM, 91191, Gif-sur-Yvette, France}
\author{Francis Bernardeau}
\affiliation{Université Paris-Saclay, CNRS, CEA, Institut de physique théorique, 91191, Gif-sur-Yvette, France}

\date{\today}

\begin{abstract}
Weak-gravitational lensing tends to wash out scale- and time-dependent features of the clustering of matter, such as the baryonic acoustic oscillations which appear in the form of wiggles in the matter power spectrum but that disappear in the analogous lensing $C_\ell$. This is a direct consequence of lensing being a projected effect. In this paper, we demonstrate how the noise complexity—often deemed "erasing the signal"—induced by a particular deprojection technique, the Bernardeau-Nishimichi-Taruya transform~\href{https://doi.org/10.1093/mnras/stu1861}{[Mon.\ Not.\ R.\ Astron.\ Soc.\ \textbf{445}, 1526 (2014)]}, can be used to extract the BAO signal and non-Gaussian aperture-mass-like properties at chosen physical scales. We take into account parts of the data vectors that should effectively be without cosmological signature and also introduce an additional reweighting designed to specifically highlight clustering features—at both the probe (summary statistics) or map (amplitude of the field) level. We thus demonstrate why weak-gravitational lensing by the large-scale structure of the Universe, though only in a tomographic setting, does not erase scale- and time-dependent features of the dynamics of matter—while providing a tool to effectively extract them from actual galaxy-shape measurements.

\end{abstract}

\keywords{Cosmology, Theory, Large-scale structures, Weak-lensing, Baryonic Acoustic Oscillations}

\maketitle

\section{Introduction}

Light rays emitted from background source galaxies propagate through the inhomogeneous distribution of baryonic and dark matter up to their observer—most likely on Earth or its L2 point given the habitat of the authors. This induces (de)magnification of the brightness of galaxies and a coherent distortion pattern in their observed shapes \citep{Bartelmann:1999yn, Kochanek:2004ua}. This effect is denoted as cosmic shear since we only detect it through the correlation of the distorted shapes of source galaxies, a consequence of the relative smallness of the effect which is hence classified as weak lensing (WL). The statistics of these WL fields thus depend on the projected three-dimensional large-scale structures (LSS) of our Universe and provide a powerful way to probe and address critical questions in cosmology such as structure formation history, the nature of dark energy and dark matter, and the laws of gravity. WL cosmology is therefore an active area of research in currently ongoing wide-area galaxy imaging surveys such as DES \citep{DES:2016jjg}, KiDS \citep{2013ExA....35...25D} and HSC-SSP \citep{2018PASJ...70S...4A}. The promising results from these surveys have also motivated the development of new generation surveys such as Euclid \citep{Euclid:2024yrr} and Vera Rubin’s LSST \citep{LSSTDarkEnergyScience:2012kar}, which will soon provide data with unprecedented quality. 

However, despite its numerous successes, WL is fundamentally limited for the study of certain LSS features given that it is by definition an integrated effect along the line of sight. This translates into scale mixing of the dynamics of matter for fixed angular scales and, thus, tends to wash out very scale- and time-dependent features in the lensing observables. One typical example is baryonic acoustic oscillations (BAOs) (see for example \cite{DESI:2025zpo} and references within), which are notably undetected in cosmic shear two-point correlations functions and only reappear in more exotic and harder to measure lensing observables such as extrema correlation functions as shown in e.g.~\cite{Gong:2025nix}. Aside from this specific example, probing the detailed scale and time evolution of the different features of the LSS is paramount to its study, maybe most notably since possible hints of dynamical dark energy appeared in the recent literature from the DESI experiment \cite{DESI:2025zgx}. This is typically done through the joint study of weak lensing and the clustering of galaxies in tomographic experiments (for example \cite{DES:2021wwk} and references within)—auto- and cross-correlations of galaxy shapes and positions—but this approach suffers from having to introduce galaxies as biased tracers of the overall matter field \cite{Desjacques:2016bnm}. 

Alternatively, sticking to cosmic shear only, the Bernardeau-Nishimichi-Taruya (BNT) transform~\cite{Bernardeau:2013rda}—a linear transformation of cosmic shear maps independent of nonlinear dynamics—has emerged in the past decade as a viable method to effectively disentangle physical scale mixing due to projection effects. However, it suffers in itself from drastically reducing the number of effective lenses which lowers the signal while at the same time suffering more from a finite density of source galaxies. In this work, we first show how to recover the signal considering the entirety of the transformed data vector, notably elements that have a null expectation value. Indeed, those still encode valuable information on the noise properties of our observables. This enables us to detect the BAO features purely in the cosmic shear signal and sets this paper within a series of works investigating alternative ways to measure BAO such as intensity mapping \cite{Rubiola:2021afc,Ostergaard:2024brd}, the Lyman-alpha forest \cite{Hadzhiyska:2025cvk,DESI:2024txa} and the use of higher-order statistics such as the bispectrum \cite{Child:2018klv,Pearson:2017wtw}. We then introduce an extra component to the BNT transform—at either the correlator or density level—demonstrating why the BAO and other scale-dependent features are not erased in the projection along the line of sight.  

This paper is organized as follows: Section~\ref{sec:BNT} recalls the basic properties and equations of the BNT transform, and Sec.~\ref{sec:detect} shows that BAO can indeed be detected in BNT-transformed weak lensing data. Section~\ref{sec:BAO} introduces a first implementation at the level of the two-point correlation function of our additional reweighting of the signal and extracts its BAO features. Section~\ref{sec:Map} presents a generalization of the argument at the map level, and Sec.~\ref{sec:non_linear} discusses different possible implementations of recipes for the nonlinear growth of cosmic structures. Section~\ref{sec:conclusion} summarizes our results and concludes.

\section{Nulling strategy}
\label{sec:BNT}

The BNT transform (sometimes referred to as nulling strategy in the earlier literature) highlights an intrinsic geometrical symmetry inherent to gravitational lensing in order to effectively select the lenses contributing to the observed effect. It thus enables one to separate the dynamical scales of matter contributing to the lensing inside the past light cone of the observer, disentangling quasilinear and nonlinear (as well as baryon-influenced) scales.

It was introduced in Ref.~\cite{Bernardeau:2013rda} and was notably used to enable very accurate from-first-principles theoretical predictions, in the context of power spectrum analysis, and of the convergence and of aperture mass probability density functions \citep{Taylor:2020zcg,Barthelemy:2020yva,Barthelemy:2023mer}. This strategy becomes even more relevant in current stage-IV lensing experiments with better knowledge of redshifts and the division of sources in more redshift bins. This has been investigated and used in various cases in recent literature \cite{Taylor:2018snp,Deshpande:2020hiu,Taylor:2020imc,Piccirilli:2025msz,Gu:2024izq}. It has also been investigated in different contexts, such as being used as a probe itself in Refs.~\cite{Touzeau:2025kdv,Touzeau:2025lwu}. And, in Ref.~\cite{Bernardeau:2020jtc}, it has been shown that the BNT transform could enable observation of the BAO in weak lensing signal. The present paper extends and generalizes their work.

We note that some other nulling strategies have been studied in the literature \cite{Huterer:2005un,Joachimi:2010va,DES:2018lpj,DES:2021jzg,Simon:2025}; however, the BNT transform appears more fundamental since it relates to symmetries in the lensing fields rather than minimizing the dependence of certain correlators to a specific effect.

In practice, the implementation of the BNT transform necessitates at least a minimum of three tomographic bins of sources. It then reduces to a linear application $p$ applied to the set of tomographic bins, which, under the Born approximation and neglecting couplings between sources (see Ref.~\cite{Bernardeau:2009bm} or Fig.~F1 of \cite{Barthelemy:2020yva} for a quantitative assessment on third-order correlations), is analogous to applying $p$ to the set of tomographic lensing efficiency kernels $\omega_i \equiv \omega(\chi|n_i(z))$ with $\chi$ the comoving distance and $n_i$ the redshift distribution of galaxies in the $i$th bin. This thus gives rise to a new set of reweighted kernels
\begin{equation}
    \BNT{\omega}_a = {p_a}^i\omega_i,
\end{equation}
where we use Einstein summation convention. For sets of three consecutive equipopulated tomographic bins arranged in ascending order, it can be shown \citep{Bernardeau:2013rda} that $p$ must satisfy the system
\begin{equation}
\left\{ \begin{aligned}
        &\sum_{i=a-2}^{a} {p_a}^i=0, \\
        &\sum_{i=a-2}^{a} \frac{{p_a}^i}{X_{i}}=0,
        \end{aligned}
\right.
\end{equation}
where
\begin{center}
\begin{equation}
    X_i = \left(\int_{z_{{\rm min}, i}}^{z_{{\rm max}, i}} {\rm d}z \, \frac{n_i(z)}{\chi(z)}\right)^{-1}
\end{equation}
\end{center}
is the inverse of the weighted average of the inverse comoving distance in the $i$th bin, analogous in spirit to a weighted average of the comoving distance in the bin.

This system of equations is underconstrained and we thus also impose by convention ${p_i}^i = 1$. The elements of $p$ can thus be computed considering sequential triplets of tomographic bins, going from the lowest to the highest redshift, such that
\begin{align}
    {p_i}^{i-2} = \frac{X_{i-2}(X_{i-1}-X_{i})}{X_{i}(X_{i-2}-X_{i-1})},\\
    {p_i}^{i-1} = \frac{X_{i-1}(X_{i}-X_{i-2})}{X_{i}(X_{i-2}-X_{i-1})}.
\end{align}

\begin{figure}
   \centering
 \includegraphics[width=\columnwidth]{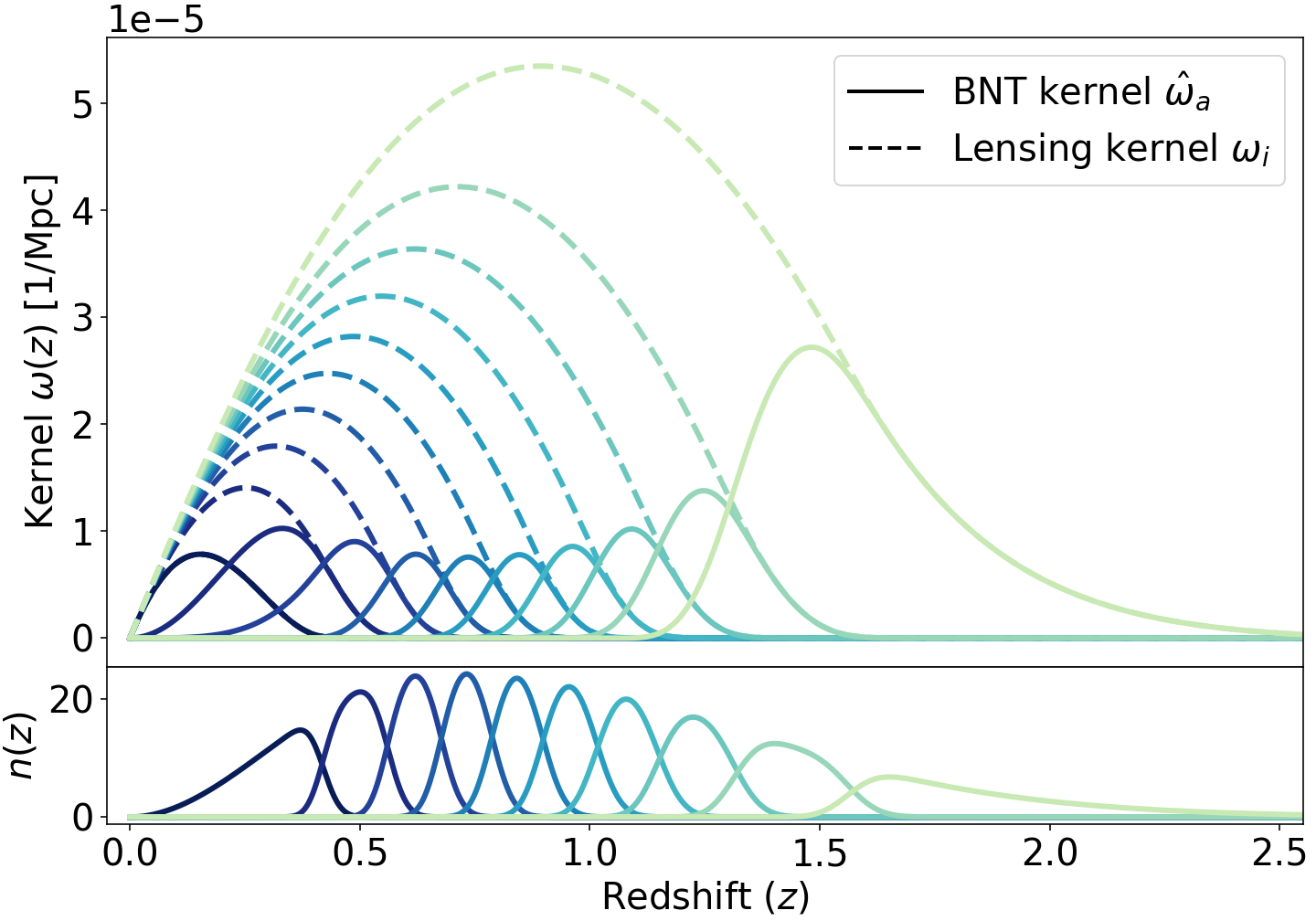}
   \caption{Lensing efficiency kernels for ten equally populated redshift bins of our semirealistic Euclid-like distribution of sources, prior to nulling (dashed lines) and after nulling (thick lines). The bottom plot shows the shape of the normalized distribution of sources inside each bin. They follow Eq.~\eqref{eq:n_iz}.}
   \label{fig:winulling}
\end{figure}

We display in Fig.~\ref{fig:winulling} a set of BNT kernels for a realistic\footnote{In this paper, we account for distribution of sources, photometric redshifts uncertainties and Gaussian shape noise in an idealized analytical framework described in Appendix~\ref{sec:euclidspec}. The chosen functionals and amplitudes are however sufficiently robust such that all our results will translate to the actual setting of the Euclid—or any other—survey.} Euclid-like distribution of sources, taking into account Euclid-like photometric redshift errors all described in Appendix~\ref{sec:euclidspec} and taken from \cite{Deshpande:2019sdl}. The displayed effect of the BNT transform is thus, as intended, to set to zero the contribution of all lenses below a chosen redshift—actually determined by the chosen tomographic binning of the sources—and therefore separating the physical scales contributing to the observed lensing effect. This property is paramount to what this paper shows, that is the ability for cosmic shear observables to probe chosen, specific physical scales.

\section{BAO detectability in weak lensing}\label{sec:detect}

\subsection{Numerical experiment}

Before any other consideration, let us show that scale-dependent features can indeed be detected in weak lensing data. To do so, we perform a Fisher analysis probing the BAO wiggles, under realistic shape noise and using the cosmic shear two-point correlation function ($\xi^+$). We first artificially split the underlying matter power spectrum in two parts: the BAO wiggles and the rest. The BAO wiggles are smoothed out following \cite{Hamann:2010pw} and the wiggles only are parametrized by their relative amplitude $A$ and their relative frequency $B$, so that our fiducial values are $A=B=1$. These two parameters in Fourier space are appropriate as they correspond in real space to the amplitude and position of the BAO peak in the two-point correlation function ($\xi^+$). In this illustrative setting, the matter power spectrum is then given by
\begin{equation}
    P_{A,B}(k)=P_{\rm nw}(k) + A[P(Bk)-P_{\rm nw}(Bk)],
\end{equation}
where $P$ is the linear matter power spectrum and $P_{\rm nw}$ its no-wiggle version, both computed using the code {\sc class} \cite{Blas:2011rf,Chudaykin:2020aoj}. We will denote by $P_{\rm w}$ the wiggle part of the matter power spectrum such that $P_{\rm w}=P-P_{\rm nw}$. Under the limber approximation, we are then able to compute the BNT-transformed two-point correlation function ($\xi^+$) in Fourier space as
\begin{equation}
   \mathcal{C}^{\BNT{\kappa} \BNT{\kappa}}_{ab}(\ell,A,B) =  \mathcal{C}^{\BNT{\kappa} \BNT{\kappa}}_{ab,{\rm nw}}(\ell) +  \mathcal{C}^{\BNT{\kappa} \BNT{\kappa}}_{ab,{\rm w}}(\ell,A,B), 
\end{equation}
where $\mathcal{C}^{\BNT{\kappa} \BNT{\kappa}}_{ab}$ denotes the BNT-transformed convergence power spectrum obtained from the BNT-transformed shear field in BNT bins $a$ and $b$ or their corresponding convergence maps $\BNT{\kappa}_a$ and $\BNT{\kappa}_b$. We also consider some realistic shape-noise contribution, which is given by $\mathcal{S}_{ij}=\delta_{ij}\sigma_{\rm s}^2/\overline{n}_i$ for the lensing power spectrum before nulling, and where $\sigma_{\rm s}=0.3$. $\overline{n}_i$ is the total number of galaxies in $i$th bin. The BNT-transformed shape noise is obtained by defining $\BNT{\mathcal{S}}_{ab}={p_a}^i \mathcal{S}_{ij} {p_b}^j$ such that the observed power spectrum is
\begin{equation}
    \observed{\mathcal{C}}^{\BNT{\kappa} \BNT{\kappa}}_{ab}(\ell) = \mathcal{C}^{\BNT{\kappa} \BNT{\kappa}}_{ab}(\ell) + \BNT{\mathcal{S}}_{ab}.
\end{equation}
For good measures, and though our result will be very weakly dependent on the knowledge of the global amplitude of shape noise, we also introduce a parameter $\alpha$, with fiducial value $1$, to the full parametrized data vector which thus reads
\begin{equation}
   \mathcal{V}_{(ab)}(\ell,A,B,\alpha) =  \mathcal{C}^{\BNT{\kappa} \BNT{\kappa}}_{ab,{\rm nw}}(\ell) +  \mathcal{C}^{\BNT{\kappa} \BNT{\kappa}}_{ab,{\rm w}}(\ell,A,B) + \alpha \BNT{\mathcal{S}}_{ab}.
\end{equation}
The fiducial covariance is assumed to follow the classical Knox approximation \cite{Knox:1995dq} and reads
{\small\begin{equation}\label{eq:covmat}
\Sigma_{(ab),(cd)}(\ell)=\frac{1}{(2\ell+1)f_{\rm sky}} \left( \observed{\mathcal{C}}^{\BNT{\kappa}\BNT{\kappa}}_{ac}(\ell)\observed{\mathcal{C}}^{\BNT{\kappa}\BNT{\kappa}}_{bd}(\ell)+\observed{\mathcal{C}}^{\BNT{\kappa}\BNT{\kappa}}_{ad}(\ell)\observed{\mathcal{C}}^{\BNT{\kappa}\BNT{\kappa}}_{bc}(\ell)\right),
\end{equation}}
where $f_{\rm sky}=0.36$, the fraction of the full sky observed by the survey, is chosen to also mimic the Euclid experiment \cite{EUCLID:2011zbd}. Our Fisher matrix reads
\begin{equation}
    F_{\theta\lambda} = \sum_{(ab),(cd),\ell} \Delta\ell \left.\frac{\partial \mathcal{V}_{(ab)}}{\partial \theta}\right|_{\rm fid}  \Sigma^{-1}_{(ab),(cd)} \left.\frac{\partial \mathcal{V}_{(cd)}}{\partial \lambda}\right|_{\rm fid}, 
\end{equation}
where $\theta$ and $\lambda$ are the parameters, $\Delta\ell$ is the size of the $\ell$ bin, and we removed the $\ell$ dependence of every term for readability.

\begin{figure*}
    \centering
    \begin{subfigure}{\columnwidth}
        \centering
        \includegraphics[width=0.99\linewidth]{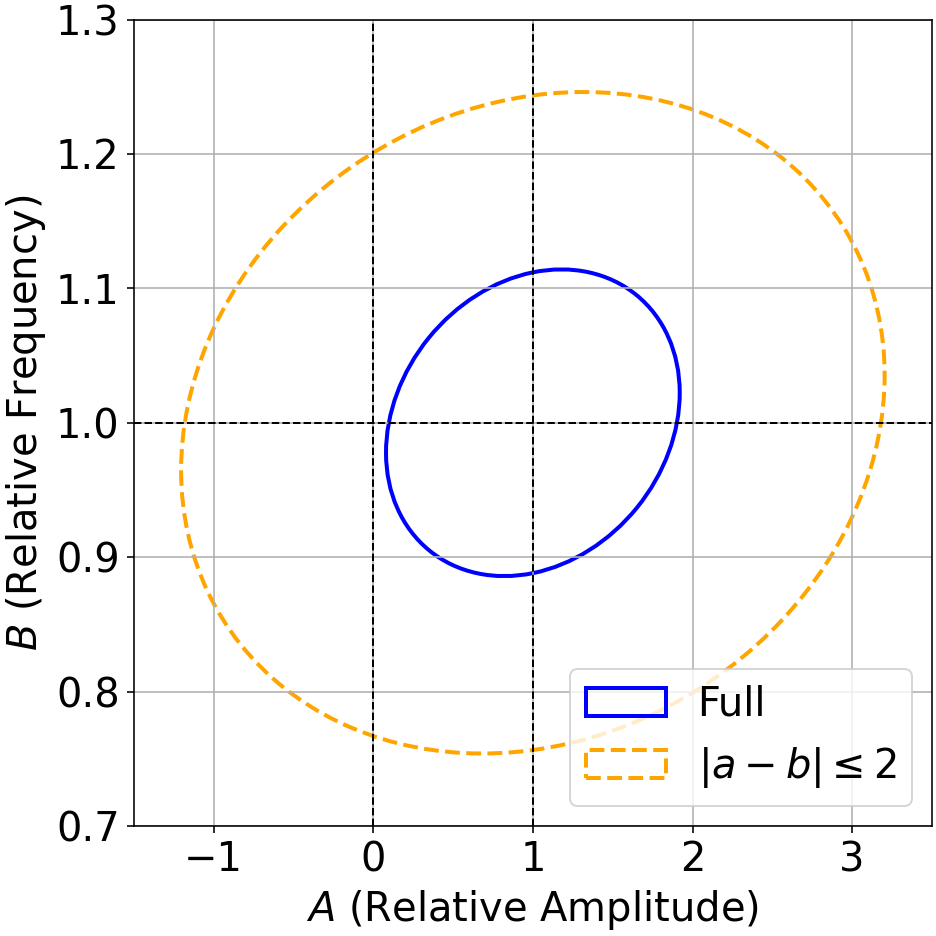}
        \caption{$30 {\rm gal/arcmin^2}$.}
        \label{fig:BAOFisher30}
    \end{subfigure}
    \hfill
    \begin{subfigure}{\columnwidth}
        \centering
        \includegraphics[width=0.99\linewidth]{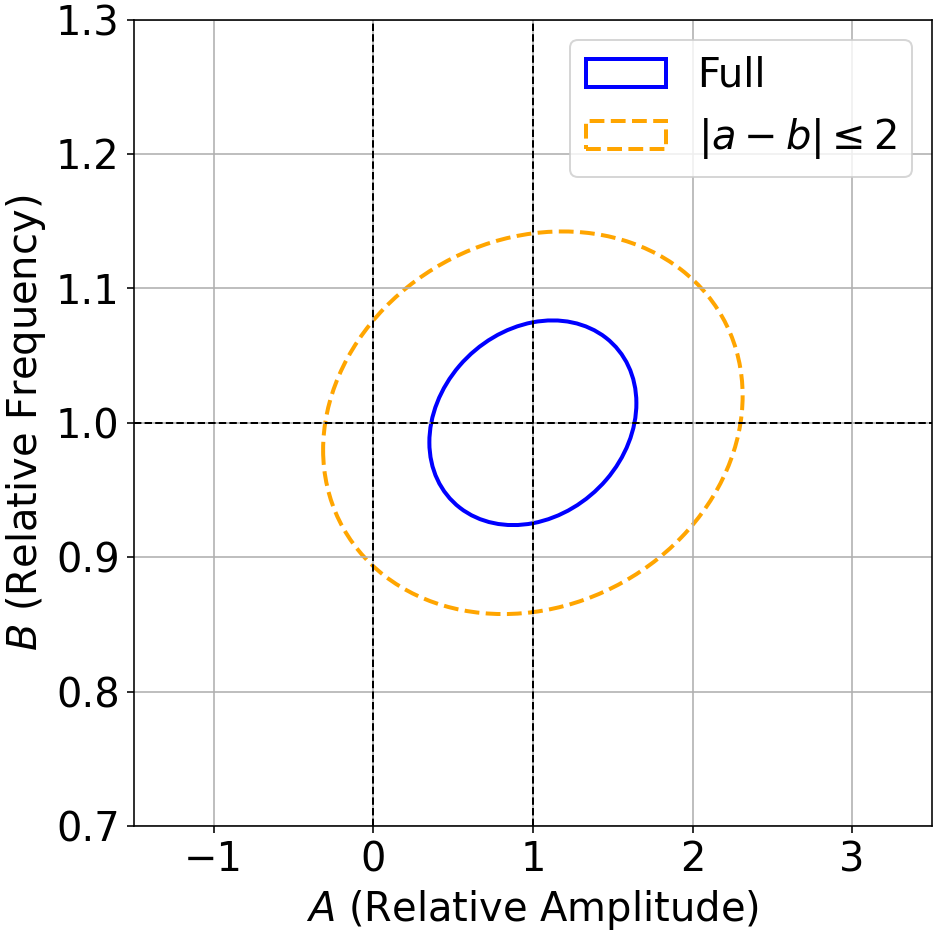}
        \caption{$60 {\rm gal/arcmin^2}$.}
        \label{fig:BAOFisher60}
    \end{subfigure}

     \caption{Constraints at $1 \sigma$ obtained with Fisher analysis for the two BAO parameters with a total galaxy number density of $30 {\rm \ gal/arcmin^2}$ and $60 {\rm \ gal/arcmin^2}$; those values can be found in Refs.~\cite{Euclid:2025plm,Euclid:2025vsf}. $A$ is the relative amplitude of the wiggles and $B$ is the relative frequency of the wiggles that is proportional to the position of the BAO peak in real space. We compare the results taking the full data vector and only correlators such that $|a-b| \leq 2$.}
   \label{fig:BAOFisher}
\end{figure*}

As an additional step and in order to make use of the full power of the implemented nulling transformation, we also perform a $k$ cut on our data vector following \cite{Taylor:2018snp}. Interested readers can check the cited paper for details, and we simply recall that given the localized kernels of the BNT transform, the $\ell$ to $k$ relationship in the harmonic two-point function is no longer ill defined. This enables a physical scale cut in our data vector, much more suited to remove uncontrolled scales (i.e. whose dynamics are not well understood) or, in this case, focus on a range of $k$ values where we would expect to see the BAO wiggles: For each correlator probing different redshifts of lenses, we only keep $\ell$ values such that their associated $k$ satisfies $4 \cdot 10^{-3} \leq k \leq 4 \cdot 10^{-1}$. Marginalizing on the noise amplitude $\alpha$ then leads to the $1\sigma$ contours in Fig.~\ref{fig:BAOFisher}.

Finally, in order to produce the contours of Fig.~\ref{fig:BAOFisher}, the choice of what to include or not in the data vector could be subject to discussion: 
On the one hand and among all pairs of $\observed{\mathcal{C}}^{\BNT{\kappa} \BNT{\kappa}}_{ab}(\ell)$, one might choose to consider only the diagonal and two subdiagonals terms, i.e. $|a-b|\leq2$, which are the only correlators whose expectation value is nonzero from the BNT transform, that is, the only correlators containing cosmological signal. This would indeed be optimal for a noiseless field, as for example considered in Ref.~\cite{Bernardeau:2020jtc} where only pairs $a=b$ were used. This choice leads to the 1$\sigma$ contour in orange dotted lines.
On the other hand, the 1$\sigma$ contour in plain blue lines displays the constraints obtained when using the full data vector (all nonredundant possible pairs of $a$ and $b$), thus including correlators whose expectation value is 0. There the figure of merit is improved by roughly a factor 4. Those (expectedly) null correlators thus clearly seem to carry extra valuable information, most notably on the noise structure which became more intricate with the BNT transform. At this stage, and though we have not yet provided an explanation for this behavior—it comes in the next subsection—we nevertheless demonstrate that the use of the BNT transform (i) enables a much better control of scale cuts and thus the validity regime of the models compared to observational sets and (ii) does not really enable a reduction the size of the data vector, even though one expects that the cosmological signature in some of its parts is null.

Figure~\ref{fig:BAOFisher} also displays how the detection of BAO features in our cosmic shear setup varies with the experiment's galaxy number density. This number influences for the most part the shape-noise contribution to cosmic shear data. We chose two particular values, $30 {\rm \ gal/arcmin^2}$ which is Euclid's requirement, currently achieved as shown in Ref.~\cite{Euclid:2025plm}, and $60 {\rm \ gal/arcmin^2}$ which could be seen a realistic value for future analysis given the performance of the Euclid Visible Camera (VIS), roughly detecting galaxy densities between 40 and 80~${\rm gal/arcmin^2}$ but rejecting around half of those due to magnitude cut and/or no reliable photometric redshift measurement (see Refs.~\cite{Euclid:2025plm,Euclid:2025vsf}).

Quantitatively speaking, with $30 {\rm \ gal/arcmin^2}$ and the full data vector, we exclude the no-wiggles scenario, $A=0$ at $1\sigma$. The detection is better with $60 {\rm \ gal/arcmin^2}$ where the no-wiggles scenario is excluded at $1.4\sigma$. Regarding the relative frequency of the wiggles $B$, whose measure is exactly equivalent to the BAO scale with fiducial value around $147 {\rm \ Mpc}$, we find a $11.4\%$ precision on $B$, at $30 {\rm \ gal/arcmin^2}$ at $1\sigma$, which immediately translates into a $11.4\%$ precision on the BAO scale. With $60 {\rm \ gal/arcmin^2}$, we find a $7.6\%$ precision at $1\sigma$. This level of precision is, of course, not competitive with current DESI \cite{DESI:2025zgx} or Planck \cite{Planck:2018nkj} whose precisions are under $1\%$, but being a completely independent estimate, with different systematics, it could improve our overall constraining power on cosmological quantities with baryonic acoustic oscillations. A Euclid-like setting is thus able to extract extra BAO information from pure cosmic shear.

\subsection{Null correlators carry valuable information: an explanation from the covariance}

An understanding of the reason why BNT correlators with null expectation values contribute so much to the amount of information extracted from the overall correlators has to do with the correlation properties among all the correlators, here identified to the covariance matrix. These correlations are the result of both shape noise (coming from looking at discrete tracers) and cosmic variance (uncertainty on the exact initial conditions) and whose structure moves from (almost) diagonal at fixed $\ell$ before the BNT transform to a more complex nontrivial form. As a consequence, one can look at the BNT transform as diagonalizing the cosmological signal covariance, identifying redundant parts in the data vector, while complexifying the noise structure associated with our lack of knowledge of the same vector. Finally, including the null correlators then allows us to learn more about the noise structure, which in the end enables better constraints on model parameters of interest.

\begin{figure*}  
  \centering
  \includegraphics[width=\textwidth]{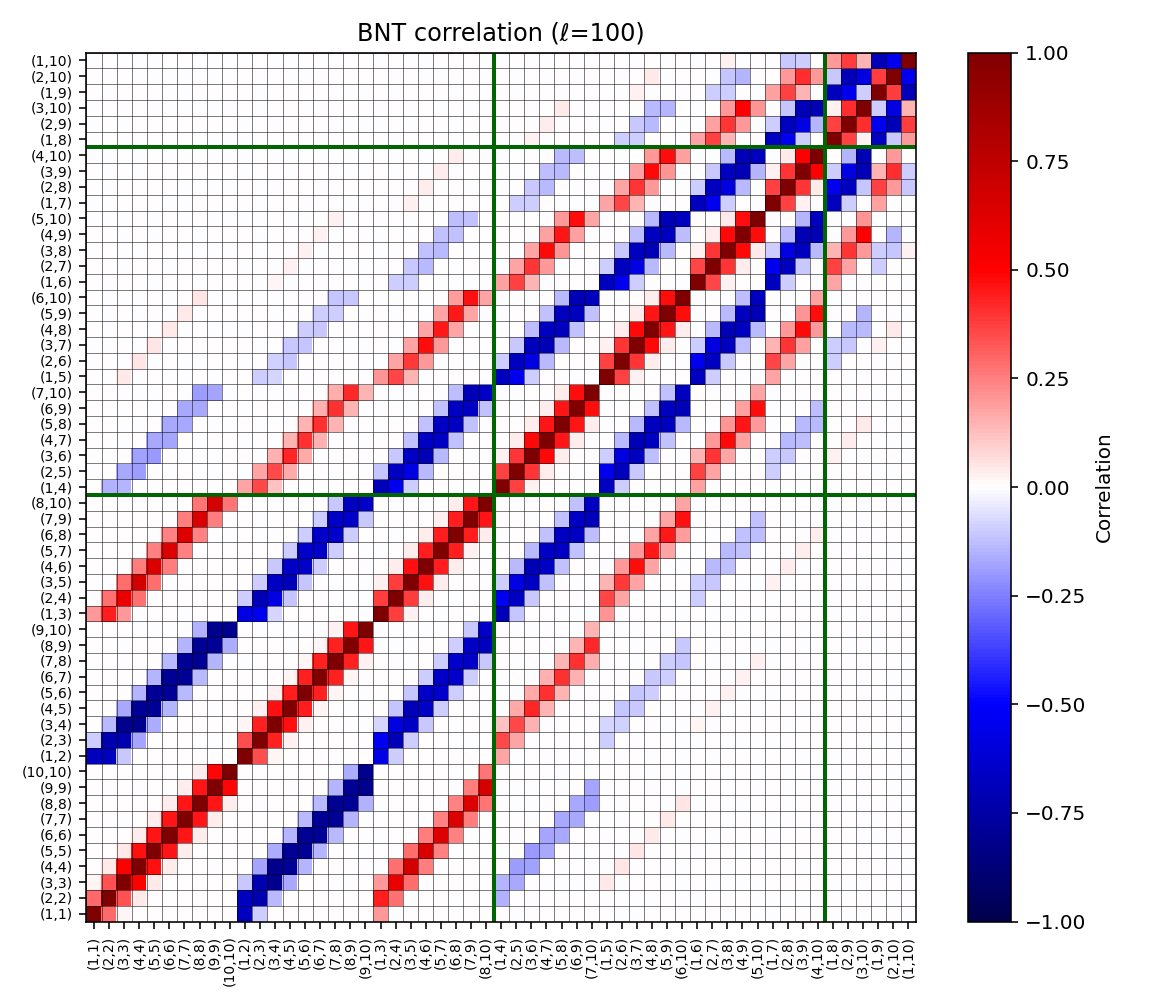}
  \caption{BNT-transformed correlation matrix at $\ell=100$. Elements are ordered per distance between indices. Green lines divide each axis between elements of the data vector of null and non-null expectation values, with two categories of null elements as described in the main text.}
  \label{fig:BNTcorr}
\end{figure*}

The correlation matrix of our BNT-transformed convergence correlators at $\ell = 100$ is shown in Fig.~\ref{fig:BNTcorr}. This choice of $\ell$ is arbitrary and for illustration purposes only; moreover, note that the matrix does not change significantly over the whole range of probed $\ell$ values. 
Qualitatively speaking, Fig.~\ref{fig:BNTcorr} clearly shows that null and non-null elements are correlated. Looking at a typical example, one can observe that $\mathcal{C}^{\BNT{\kappa}\BNT{\kappa}}_{48}$ is correlated with $\mathcal{C}^{\BNT{\kappa}\BNT{\kappa}}_{66}$, $\mathcal{C}^{\BNT{\kappa}\BNT{\kappa}}_{67}$, $\mathcal{C}^{\BNT{\kappa}\BNT{\kappa}}_{78}$, $\mathcal{C}^{\BNT{\kappa}\BNT{\kappa}}_{46}$,
$\mathcal{C}^{\BNT{\kappa}\BNT{\kappa}}_{57}$ and $\mathcal{C}^{\BNT{\kappa}\BNT{\kappa}}_{68}$.\footnote{As an example, from Eq.~\eqref{eq:covmat}, note that in the Gaussian approximation we have $\left\langle \mathcal{C}^{\BNT{\kappa}\BNT{\kappa}}_{48} \mathcal{C}^{\BNT{\kappa}\BNT{\kappa}}_{66}\right\rangle (\ell) = \frac{2}{(2 \ell +1) f_{\rm sky}}\observed{\mathcal{C}}^{\BNT{\kappa}\BNT{\kappa}}_{46} (\ell) \observed{\mathcal{C}}^{\BNT{\kappa}\BNT{\kappa}}_{68} (\ell)$. Both $\mathcal{C}^{\BNT{\kappa}\BNT{\kappa}}_{46}$ and $\mathcal{C}^{\BNT{\kappa}\BNT{\kappa}}_{68}$ expectation values are nonvanishing after the BNT transform, leading to a nonzero element of the covariance matrix. Moreover $\observed{\mathcal{C}}^{\BNT{\kappa}\BNT{\kappa}}_{46}$ and $\observed{\mathcal{C}}^{\BNT{\kappa}\BNT{\kappa}}_{68}$, respectively, include $\observed{\mathcal{C}}^{\kappa\kappa}_{44}$ and $\observed{\mathcal{C}}^{\kappa\kappa}_{66}$ which implies the presence of shape noises $\mathcal{S}_{44}$ and $\mathcal{S}_{66}$ that appear to be the dominant terms for this element, as well as most of the others, of the covariance.} More specifically, ordering indices in the covariance matrix of Fig.~\ref{fig:BNTcorr} according to $a-b$ of $\mathcal{C}^{\BNT{\kappa}\BNT{\kappa}}_{ab}$, we divide each axis of the matrix into three parts with green lines: These define three categories of correlators which in turn enable comparison of the correlations of each category with the others. For $\BNT{\mathcal{C}}^{\BNT{\kappa}\BNT{\kappa}}_{ab}$, the three categories correspond to $|a-b|\leq2$ which are non-null correlators together, $2<|a-b|\leq6$ which are null correlators that have significant correlations with non-null correlators, and $6<|a-b|$ which are null correlators that are uncorrelated with non-null correlators but are correlated with the previous category of null correlators. Technically, only the first category contains cosmological information. The second set contains cosmic-variance and shape-noise information that is correlated with the data from the first set. The third set enables refining cosmic-variance and shape-noise information due to its correlation with the second set. Given this structure, and strictly mathematically speaking given that the matrix is not diagonal per block, nothing can be thrown away without loss of information and thus loss in constraining power of model parameters. To say the same thing in a different fashion, the inverse of this covariance matrix which will in the end translate into error bars in parameter space is influenced by every block defined by the green lines in Fig.~\ref{fig:BNTcorr}, blocks which cannot be inverted independently of the others. This is indeed consistent with what we found in the previous subsection in our numerical experiment.

\section{Recovering BAOs in lensing: Probe-level reconstruction}
\label{sec:BAO}

The previous section demonstrated how BNT-transformed two-point correlators can be used to estimate the BAO in pure lensing datasets. While we previously insisted on the actual methodology to extract those features from realistic data with different noise contributions, we now turn to the pure cosmological signal and demonstrate where, in the mathematical structure of the lensing fields, we can extract scale- and time-dependent features in the field of the matter density contrast. This section will serve as a recap of previous results obtained in Ref.~\cite{Bernardeau:2020jtc} on the specific reconstruction of BAOs in the lensing power spectra. The next section (Sec.~\ref{sec:Map}) consistently generalizes this approach at the field level and Sec.~\ref{sec:non_linear} comments on aspects of the modeling of nonlinear dynamics of matter, mainly independent from the geometric arguments we want to emphasize but that could reveal important to recovering the cosmological features in lensing surveys.

\subsection{BAO wiggles in the cosmic shear power spectra}

\begin{figure}
    \centering
    \begin{subfigure}{\columnwidth}
        \centering
        \includegraphics[width=0.99\linewidth]{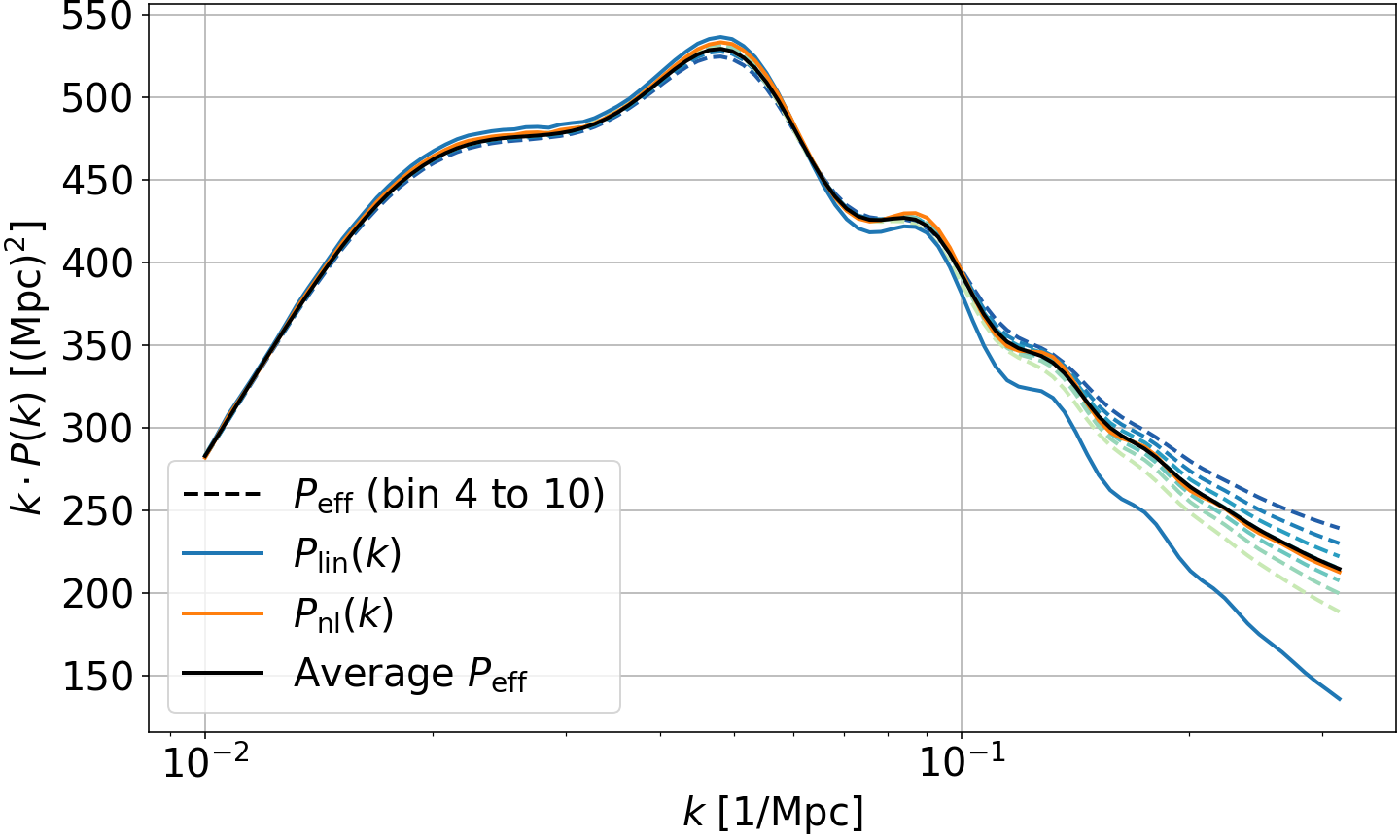}
        \caption{Rescaled effective power spectrum $k \cdot P_{\mathrm{eff}}(k)$ at $z=1$ for each bin and their average.}
        \label{fig:Pefflin}
    \end{subfigure}

    \begin{subfigure}{\columnwidth}
        \centering
        \includegraphics[width=0.99\linewidth]{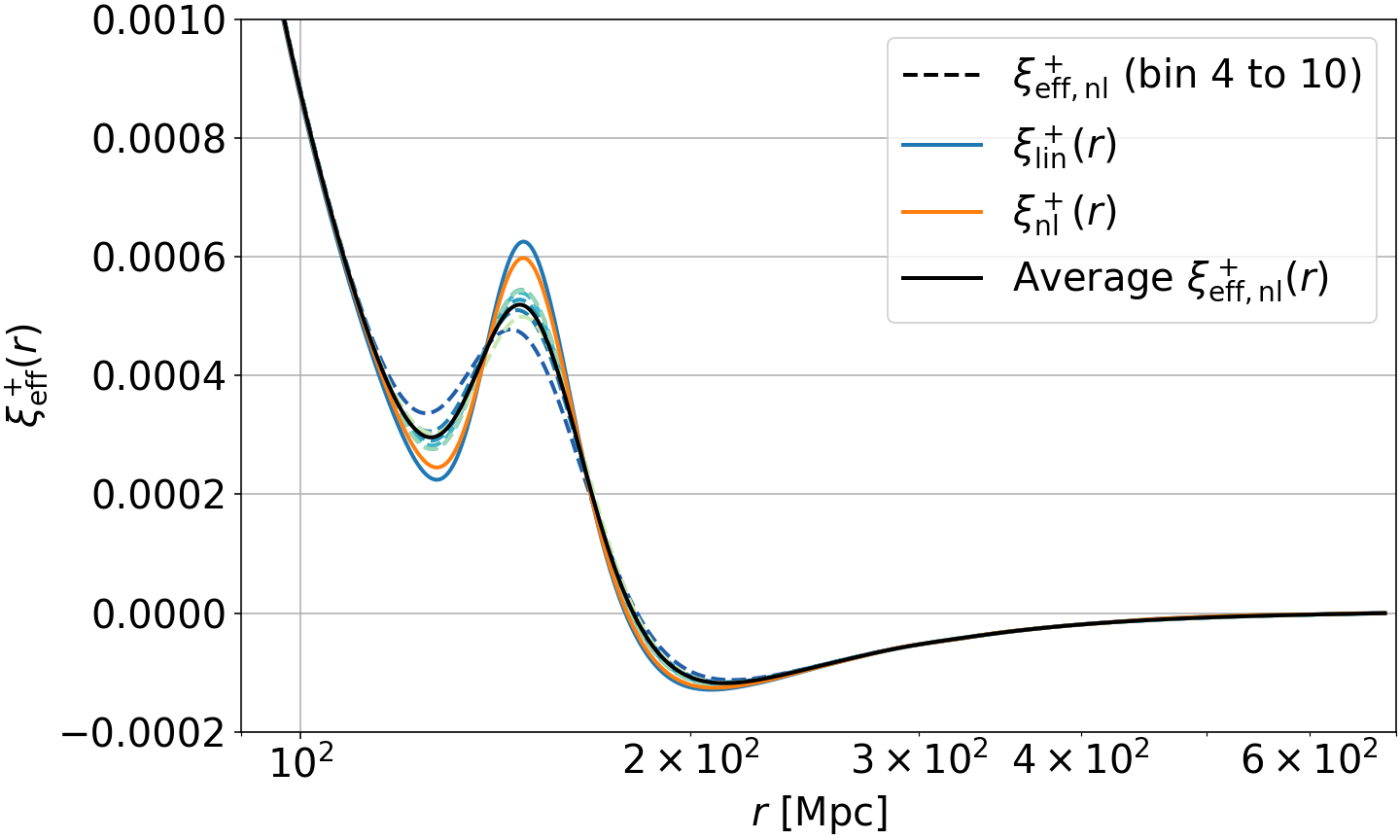}
        \caption{Effective two-point correlation function $\xi^+_{\mathrm{eff}}(r)$ reconstructed from $P_{\mathrm{eff}}$.}
        \label{fig:xiefflin}
    \end{subfigure}

    \caption{Comparison of probe-level effective power spectra and their real-space counterpart from auto bin lensing correlations.}
    \label{fig:Peff_xi_combined}
\end{figure}

From the application of the BNT transform to the different tomographic bins, the cosmic shear autopower spectrum of bin $a$ is given by  
\begin{equation}
   \mathcal{C}^{\BNT{\kappa} \BNT{\kappa}}_{aa}(\ell) = \int \frac{\dd \chi}{\chi^2} \BNT{\omega}_a^2(\chi)P\left(\frac{\ell}{\chi},z(\chi)\right),
\end{equation}
where $P$ is the matter power spectrum. If one were to be interested in BAO wiggles, it might be reasonable to consider that the scales of interest ($k \sim 0.05$) could as a first approach be considered to evolve with respect to linear theory. One can then formally decouple scale and time evolution such that the auto shear power spectra now take the form
\begin{equation}
   \mathcal{C}^{\BNT{\kappa} \BNT{\kappa}}_{aa}(\ell) = \int \frac{\dd \chi}{\chi^2} \BNT{\omega}_a^2(\chi)\frac{D_+^2(\chi)}{D_+^2(z_{\rm ref})} P\left(\frac{\ell}{\chi},z_{\rm ref}\right).
\end{equation}
Here $D_+$ is the linear growth rate of the matter fluctuations normalized such that $D_+(z=0)=1$ and we postpone the discussion on nonlinear dynamics to Sec.~\ref{sec:non_linear}. Then, the comoving distance  of the most contributing lens to the power spectrum in bin $a$ is given by $\langle \chi \rangle_{aa}$ for which the BNT lensing autopower spectra trace the matter power spectrum at $ k = l/\langle\chi\rangle_{aa}$:
\begin{equation}
    \left< \chi \right>_{aa} = \frac{\int \frac{\dd \chi}{\chi^2} \chi \BNT{\omega}_a^2(\chi)}{ \int \frac{\dd \chi}{\chi^2} \BNT{\omega}_a^2(\chi)}.
\end{equation}
Note that we denote this distance with an expectation value $\langle \ \rangle$ since it would coincide with a more generic definition of the mean lens distance within bin $a$ the narrower the bin becomes through the BNT transform.\footnote{In that case, the $\BNT{\omega}_a^2/\chi^2$ term in the integrand would be simply replaced by $\BNT{\omega}_a$.} This is what we will do for a more generic recovery of the matter density contrast in Sec.~\ref{sec:Map}. Finally, relying on the stationary phase approximation (since the BNT narrow lensing kernel varies rapidly compared to the matter power spectrum), we define an effective lens efficiency\footnote{The choice to perform the growth factor rescaling of the linear matter power spectra within the line-of-sight integral [Eq.~\eqref{eq:Weffprobe}] or outside after the geometrical re-scaling [Eq.~\eqref{eq:Peffprobe}] is \textit{a priori} arbitrary. It however turns out that, from a pure functional point of view, the stationary phase approximation is (very slightly) better behaved in the case where only the linear $P(\ell/\chi,z_{\rm ref})$ is taken outside of the integral.}
\begin{equation}
    W_{\text{eff},aa}=\int \frac{\dd \chi}{\chi^2} \BNT{\omega}_a^2(\chi)\frac{D_+^2(\chi)}{D_+^2(z_{\rm ref})}
    \label{eq:Weffprobe}
\end{equation}
such that we can recover an effective matter power spectrum
\begin{equation}\label{eq:Peffprobe}
    P_{\text{eff},a}(k)=\frac{\mathcal{C}^{\BNT{\kappa} \BNT{\kappa}}_{aa}(k \left< \chi \right>_{aa}) }{W_{\text{eff},aa}} \approx P(k,z_{\rm ref}) \ {\rm at \ chosen \ }z_{\rm ref}, 
\end{equation}
displayed in Fig.~\ref{fig:Pefflin} for $z_{\rm ref} = 1$.
In this figure, one can observe the BAO wiggles of the matter power spectrum directly from the cosmic shear which illustrates our ability to probe scale- and time-dependent features despite projection effects. Note however that the wiggles are still damped as the lensing kernels are not infinitely thin, a direct consequence of having a finite density of sources on the observed sky. One can expect a monotonic evolution of this damping with the thinness of the BNT-transformed weak lensing kernels, an effect we quantify and discuss in Appendix~\ref{sec:tomoprec}. Moreover, uncontrolled photometric redshift errors—meaning beyond Gaussian errors that we consider here—such as catastrophic outliers that could appear in a real experimental setup could potentially become an important systematic to be taken into account. We discuss this potential issue in Appendix~\ref{sec:photo_error}.

Aside from the slightly damped wiggles, the average effective $P_{\rm eff}(k)$, in black, reconstructed from the lensing power spectra, matches the nonlinear power spectra to very good agreement (a few percent in the worst cases), even assuming linear evolution for the reconstruction. We refer the reader to Fig.~\ref{fig:Peff_xi_nl_combined} for the equivalent plot taking into account the nonlinear evolution of the underlying matter power spectrum in the reconstruction. At the level of the reconstructed matter power spectra coming from each redshift bin, they oscillate around the nonlinear curve for scales where nonlinearities become dominant in the matter power spectrum ($k \gtrsim 0.1$). Indeed, each lensing power spectrum in their respective lensing BNT bin geometrically traces the matter power spectrum at a redshift corresponding to the comoving distance of the most contributing lens in the bin. Then, whatever dynamical evolution is assumed, here linear theory, the rescaling through $W_{\rm eff}$ will "evolve" the observable to a common redshift of interest, here $z = 1$ and lenses narrowly centered around $z = 1$ are purely geometrically rescaled with no assumption on the lens dynamical evolution. This particular effective power spectra will then closely match the theoretical pure nonlinear spectrum at $z = 1$, and contributions from other bins will oscillate around this curve, in a manner dependent on their most contributing lenses (whether below or above $z = 1$) and the model chosen for the dynamical evolution.

Finally, note that a similar construction could be performed using the lensing cross-spectra $\mathcal{C}_{ab}$ where $a \neq b$, still defining a mean comoving distance with two (overlapping) kernels instead of one squared. We do not include those in the above plots since their signal is damped with respect to the auto spectra and thus mostly contribute to noise reduction as seen in the previous section.

\subsection{BAO in the lensing two-point correlation function}

Cosmic shear experiments tend to rather measure the analogous two-point shear correlation function, which we here study through its component $\xi^+$, the real counterpart of the convergence power spectrum. Through linearity of the inverse Fourier transform, and assuming as in the previous subsection linear evolution of the underlying power spectrum, the exact same construction, reweighting the nulled $\hat{\xi}^+(\theta)$ by $W_{\text{eff},aa}$ and taking $r = \langle \chi \rangle_{aa} \times \theta$, enables one to probe the two-point correlation function of the matter field. From the already constructed $P_{{\rm eff},a}(k)$, we have
\begin{equation}
    \xi^+_{\text{eff},a}(r)=\int dk \frac{k^2}{2 \pi^2} {\rm sinc}(kr) P_{\text{eff},a}(k),
\end{equation}
which we display in Fig.~\ref{fig:xiefflin}. There, the BAO peak is measured at around $147 {\rm \ Mpc}$, similarly demonstrating the ability of our proposed implementation to reveal scale-dependent features in the cosmic shear. However, the peak amplitude is a little damped in the reconstructed correlation function with respect to its 3D counterpart, which is another presentation of the damping of the wiggles in Fourier space, still due to nonzero width of BNT-transformed lensing kernels. That said, the width of the chosen bins are those of a realistic Euclid-like survey, leading to the ability of Euclid to detect BAO in lensing two-point observables. This is in agreement with our previous results in Sec.~\ref{sec:detect}.

\section{Field-level reconstruction}
\label{sec:Map}

In this section, we extend the probe-level reconstruction technique of the previous section to the field level so that it can be used for various observables. As an example, we will write a systematic inversion equation for every high-order cumulant of the matter density contrast. 

\subsection{(Naïve) field-level inversion}

The lensing convergence smoothed within a symmetric window of angular size $\theta$ after BNT transform can be written as:
\begin{equation}
    \BNT{\kappa}_a = \int \dd \chi \BNT{\omega}_a(\chi) \delta_m(\theta \chi,z_\chi),
\end{equation}
where $\delta_m(\theta \chi,z_\chi)$ is the matter density contrast smoothed at scale $\theta \chi$ and taken at the redshift corresponding to $\chi$.
Similarly to the two-point function in the previous section, the stationary phase approximation enables one to define a field-level effective kernel
\begin{equation}
    w_{\text{eff},a}= \int \dd \chi \BNT{\omega}_a(\chi)
\end{equation}
such that we obtain
\begin{equation}
    \BNT{\kappa}_a \approx  w_{\text{eff},a} \times \delta_m(\theta \chi_{{\rm eff},a}, z_{\chi_{{\rm eff},a}}), 
    \label{eq:map_kappa}
\end{equation}
where $\chi_{{\rm eff},a}$, whose definition is discussed in the next subsection, is the effective comoving distance of the lenses contributing to the $a$th BNT-transformed bin. This in principle enables the reconstruction of different correlators of the density contrast, with a single effective kernel.

\subsection{Correlators in the Limber approximation}
\label{ssec:saddlepoint}

Ideally and from Eq.~\eqref{eq:map_kappa}, one could want to recover any cumulant (or more generally any correlators) of the matter density contrast $\left\langle\delta_m^n\right\rangle_c$ in "2D" (see footnote~\ref{foot5}) from $\left\langle\BNT{\kappa}_a^n\right\rangle_c$ through a rescaling from $w_{\text{eff},a}^n$. However, and from a simple dimensional analysis, this cannot be correct and some extra geometric factor would need to be taken into account, as could already be seen at the two-point level where the Limber approximation commanded an extra $1/\chi^2$ factor in its specific effective kernel $W_{\rm eff}$.

\begin{figure}
    \centering

    \begin{subfigure}{\columnwidth}
        \centering
        \includegraphics[width=0.99\linewidth]{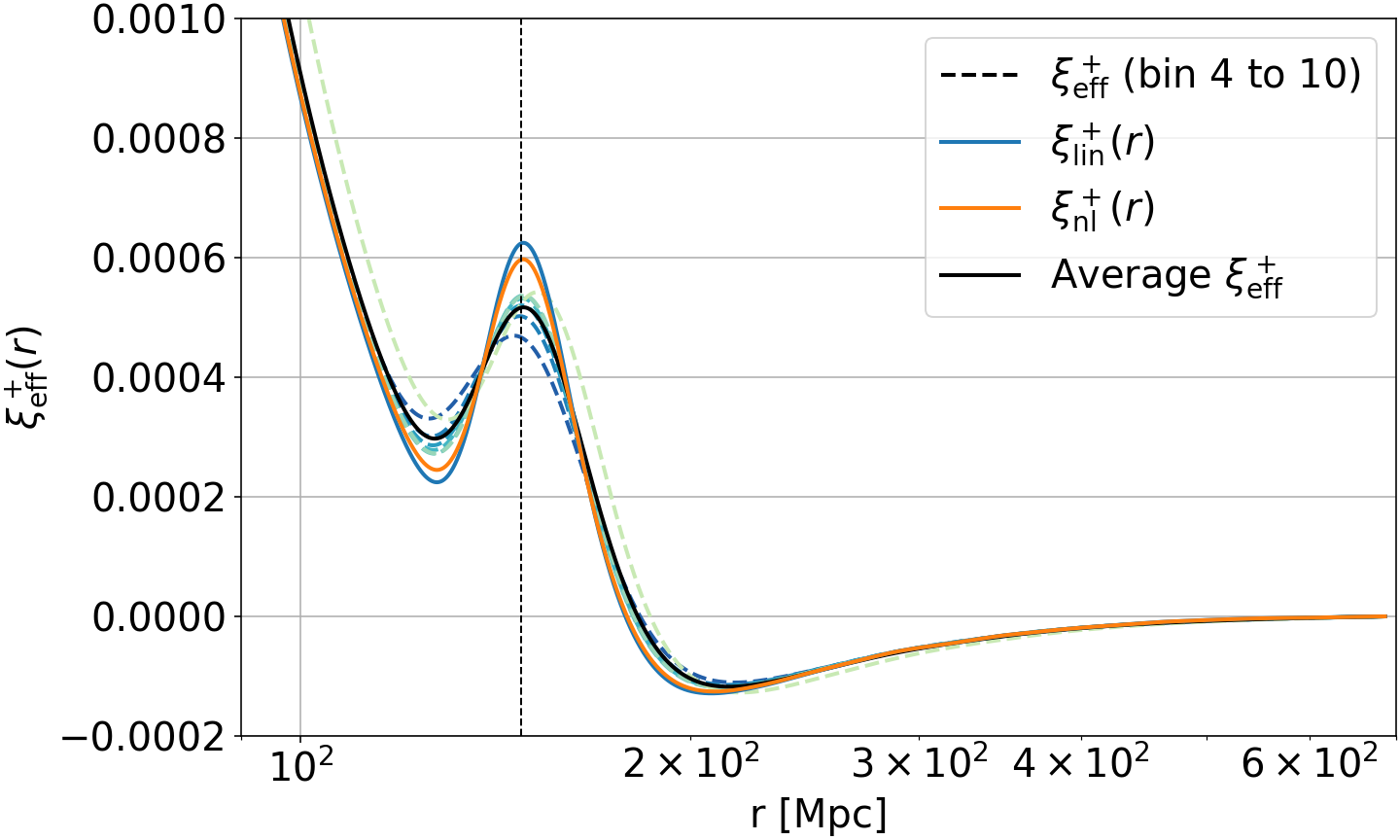}
        \caption{Effective two-point correlation function $\xi^+_{\mathrm{eff}}(r)$ computed using the \textit{geometric} definition.}
        \label{fig:xieff_dist}
    \end{subfigure}

    \begin{subfigure}{\columnwidth}
        \centering
        \includegraphics[width=0.99\linewidth]{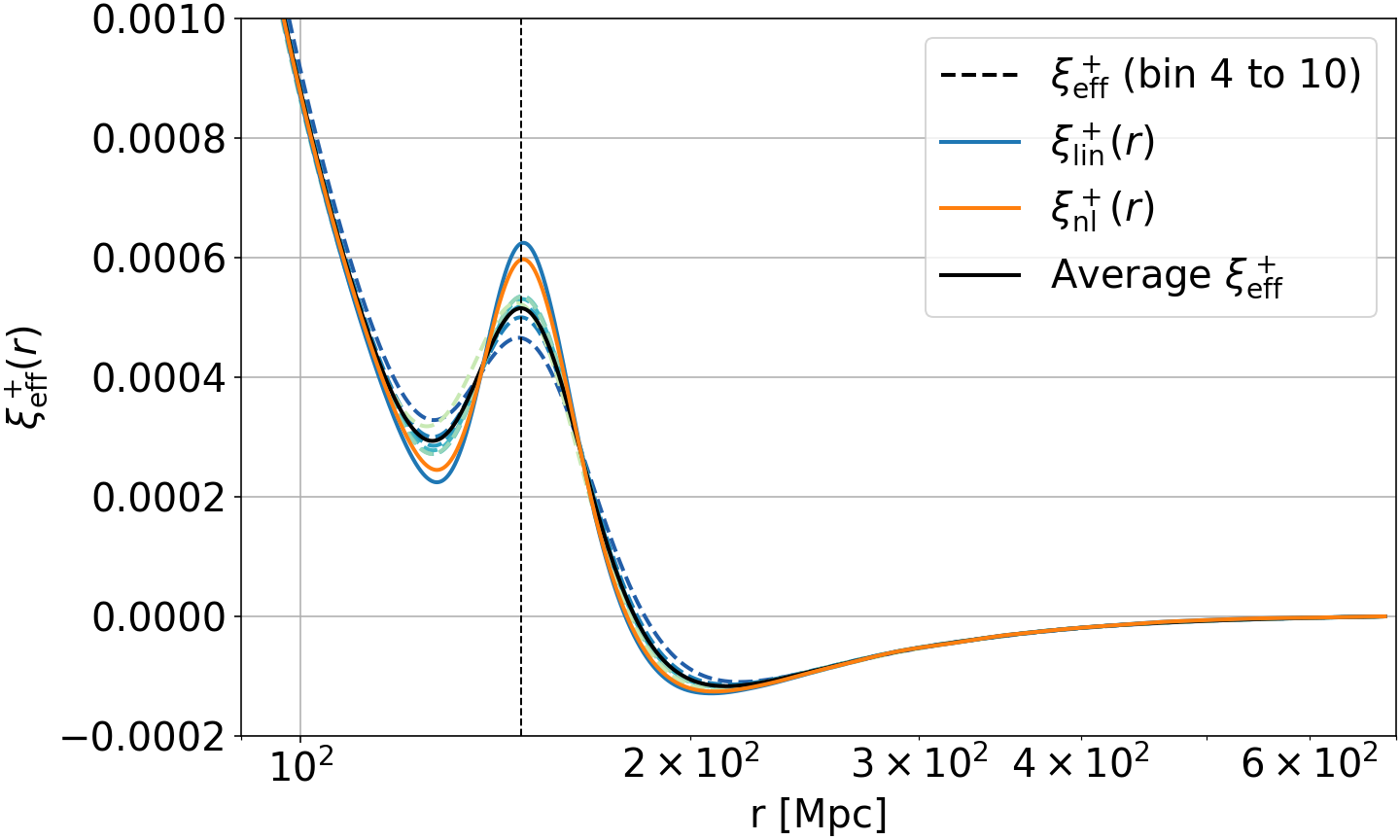}
        \caption{Effective two-point correlation function $\xi^+_{\mathrm{eff}}(r)$ computed using the \textit{saddle-point} definition.}
        \label{fig:xieff_deriv}
    \end{subfigure}

    \caption{Comparison of real-space effective correlation functions reconstructed using the two different definitions for $\chi_{{\rm eff},a}$ and $\sigma^2_{\chi,a}$ in Eq.~\eqref{eq:omega_bnt_eff}. Each plot shows bin autocorrelations and their average, along with the matter two-point function in linear and nonlinear theory.}
    \label{fig:xi_eff_maplvl_compare}
\end{figure}

In order to systematically compute the needed geometrical factor, note that following the Limber approximation we have
\begin{equation}\label{eq:cumulantlimber}
    \left\langle\BNT{\kappa}_a^n \right\rangle_c=\int \dd \chi \BNT{\omega}_a^n(\chi) \left\langle\delta_m^n (\theta \chi,z_\chi)\right\rangle_c,
\end{equation}
where the BNT lensing kernel $\BNT{\omega}_a$ is narrow enough so that we can perform a leading-order saddle-point approximation controlled by the exponent $n$. We finally obtain
\begin{equation}\label{eq:cumulant}
    \left\langle\delta_m^n(\chi_{{\rm eff},a}\theta,z_{\chi_{{\rm eff},a}})\right\rangle_c \approx \frac{\left\langle\BNT{\kappa}_a^n\right\rangle_c}{\left(\BNT{\omega}_a(\chi_{\text{eff},a})\right)^n} \times \sqrt{\frac{n}{2 \pi \sigma_{\chi,a}^2}},
\end{equation}
which is exact in the limit where $n$ goes to infinity.
In the previous equation, the mean $\chi_{\text{eff},a}$ and the standard deviation $\sigma_{\chi,a}$ are both obtained from reading the BNT lensing kernel as a normal distribution whose cumulants can be computed either in the pure saddle-point fashion, that is, Taylor expanding $\exp[n\, \ln(\BNT{\omega}_a)]$ around its maximum where the gradient is zero, or estimating a geometric mean and variance. In both cases we have
\begin{equation}\label{eq:omega_bnt_eff}
    \BNT{\omega}_a(\chi) \approx \BNT{\omega}_a(\chi_{\text{eff},a}) \times \exp(-\frac{(\chi-\chi_{\text{eff},a})^2}{2\sigma_{\chi,a}^2}),
\end{equation}
where $\chi_{\text{eff},a}$ and $\sigma_{\chi,a}$ are defined in Table~\ref{tab:chieff}.
\begin{table}[]
\caption{\label{tab:chieff}Definition of $\chi_{\text{eff},a}$ and $\sigma_{\chi,a}$ in Eq.~\eqref{eq:omega_bnt_eff} from their "geometric" or "saddle-point" definition.}
\begin{ruledtabular}
\begin{tabular}{ScScSc}
\ \  \ \ \ \ \ \ \ \ \ \ \ & \, Geometric definition \, & \, Saddle-point definition \, \\
\hline
$\chi_{\text{eff},a}$ & $ \bar{\chi} = \dfrac{\int \dd\chi \chi \BNT{\omega}_a}{\int \dd\chi \BNT{\omega}_a}$ & 
 $\arg\max_\chi \BNT{\omega}_a(\chi)$ \\
$\sigma_{\chi,a}^2$ &
$\dfrac{\int \dd\chi \chi^2 \BNT{\omega}_a}{\int \dd\chi \BNT{\omega}_a} - \bar{\chi}^2$ &
$-\left( \dfrac{\BNT{\omega}_a(\chi_{\text{eff},a})}{\BNT{\omega}_a''(\chi_{\text{eff},a})} \right)$ \\
\end{tabular}
\end{ruledtabular}
\end{table}
Note that from a purely mathematical point of view, the saddle-point definition will better probe the maximally contributing physical scales. We exemplify this last point in Fig.~\ref{fig:xi_eff_maplvl_compare} where we again reconstruct the matter two-point correlation function from its lensing counterpart in different source bins, this time using the generalized reconstruction method of this section. Although very similar, the two plots—each taking different definitions for $\chi_{{\rm eff},a}$ and $\sigma^2_{\chi,a}$—exhibit tiny differences for each tomographic bin to the advantage of the saddle-point definition for $\chi_{{\rm eff},a}$ and $\sigma^2_{\chi,a}$ since the BAO peak is there more consistently localized. 

In the case of $n$-point spatial correlators in Fourier space, Eq.~\eqref{eq:cumulantlimber} is modified such that $\BNT{\omega}^n \rightarrow \BNT{\omega}_{a,b,...}^n/\chi^{2n-2}$, where $\BNT{\omega}_{a,b,...}^n$ represents the product of $n$ kernels of the considered source bins. This does not change our reconstruction through the saddle-point method. Moreover, since the maximum argument of $\BNT{\omega}^n$ and $\BNT{\omega}^n/\chi^{2n-2}$ are extremely similar, and even exactly the same in the infinitely narrow kernel approximation, we keep the same $\chi_{{\rm eff}}$ and $\sigma^2_{\chi}$ such that we obtain
\begin{equation}\label{eq:Peffmap}
    P_{\text{eff},a}(k)=\frac{\chi_{\text{eff},a}^2}{\sqrt{\pi \sigma_{\chi,a}^2}} \frac{\mathcal{C}^{\BNT{\kappa} \BNT{\kappa}}_{aa}(k \chi_{\text{eff},a}) }{\left(\BNT{\omega}_a(\chi_{\text{eff},a})\right)^2} \frac{D_+^2(z_{\rm ref})}{D_+^2(z_{\chi_{{\rm eff},a}})} \approx P(k,z_{\rm ref}),
\end{equation}
where the different terms compared to Eq.~\eqref{eq:Peffprobe} explicit the volume of the long cylinder in which one computes the matter power spectrum before projection,\footnote{Rigorously formulated, the Limber approximation for projected quantities does not add-up statistically independent 2D slices of matter but rather statistically independent long cylinders centered on the matter slices. See for example the demonstration given in section 2.3 of \cite{2020MNRAS.492.3420B}.\label{foot5}} as well as the assumed dynamical evolution of the matter power spectrum.

We use the saddle-point definitions for the reminder of the paper.

\subsection{Cumulant Generating Function(al)}

From Eq.~\eqref{eq:cumulant}, the cumulant generating function of the matter density contrast $\delta_m$ from the cumulants of the BNT-transformed convergence $\BNT{\kappa}_a$ reads
\begin{align}
&\phi_{\delta_m}(\lambda)|_{z = z_{\rm ref}}=\sum_{n=1}^{+\infty}\frac{\left\langle\delta_m^n\right\rangle_c}{n!} \lambda^n \\
\approx &\sum_{n=1}^{+\infty} \frac{\left\langle\BNT{\kappa}_a^n\right\rangle_c}{n!} \times \sqrt{\frac{n}{2 \pi \sigma_{\chi,a}^2}} \left(\frac{\lambda}{\BNT{\omega}_a(\chi_{\text{eff},a})}\right)^n \times \frac{D_+^{2n-2}(z_{\rm ref})}{D_+^{2n-2}(z_{\chi_{{\rm eff},a}})},
\end{align}
where we have omitted the smoothing scale $\chi_{{\rm eff},a} \theta$ for clarity and where the scaling with the linear growth factor follows tree-order Eulerian perturbation theory.
Note that a similar construction holds for correlators of rank $n$ (such as the power spectrum at $n =2$, the bispectrum at $n=3$, etc), as illustrated in Eq.~\eqref{eq:Peffmap}, since the geometric factors are the same as for cumulants of rank $n$. This enables writing the cumulant generating functional of the matter density field which generates every correlators. In the case of autocorrelators in the same bin we indeed obtain
\begin{align}
    \label{eq:functional1}
    {\rm ln}\mathcal{Z}_{\delta_m}(\mathbf{J}) = &\sum_{n=1}^{\infty}\frac{1}{n!}\int\frac{{\rm d}^2k_1}{(2\pi)^2}\cdots\int\frac{{\rm d}^2k_n}{(2\pi)^2}\nonumber\\
    &\times\langle\tilde{\delta}_m(\mathbf{k}_1)\cdots\tilde{\delta}_m(\mathbf{k}_n)\rangle_{c} \, \mathbf{J}(\mathbf{k}_1)\cdots\mathbf{J}(\mathbf{k}_n) 
\end{align}
with
\begin{align}
    \label{eq:functional2}
    \langle\tilde{\delta}_m(\mathbf{k}_1)\cdots\tilde{\delta}_m(\mathbf{k}_n)\rangle_{c} \approx & \, \sqrt{\frac{n}{2 \pi \sigma_{\chi,a}^2}}\frac{\chi_{{\rm eff},a}^{2n-2}}{\BNT{\omega}^n_a(\chi_{\text{eff},a})} \frac{D_+^{2n-2}(z_{\rm ref})}{D_+^{2n-2}(z_{\chi_{{\rm eff},a}})} \nonumber\\
    &\times \langle\tilde{\kappa}_a(\mathbf{k}_1\chi_{{\rm eff},a})\cdots\tilde{\kappa}_a(\mathbf{k}_n\chi_{{\rm eff},a})\rangle_{c}.
\end{align}
Additionally, one could imagine taking angular derivatives of the BNT-transformed convergence which would translate into spatial derivatives of the matter density contrast. In that sense, and since most geometrical (e.g. peaks, voids, extrema in general etc.) and topological (e.g. minkowski functionals) quantities of the field can be expressed from series of generalized cumulants of the amplitude of the field and its derivatives (see for example \cite{gay}), one could imagine probing such features of the (cylindrically) smoothed matter density contrast. The quantitative exploration of these aspects is left for future work but all the geometric ingredients related to the BNT transform presented in this work would suffice for at least formal definitions.

\subsection{An illustration: The matter density skewness from the lensing convergence}

\begin{figure}
    \centering
    \includegraphics[width=0.99\linewidth]{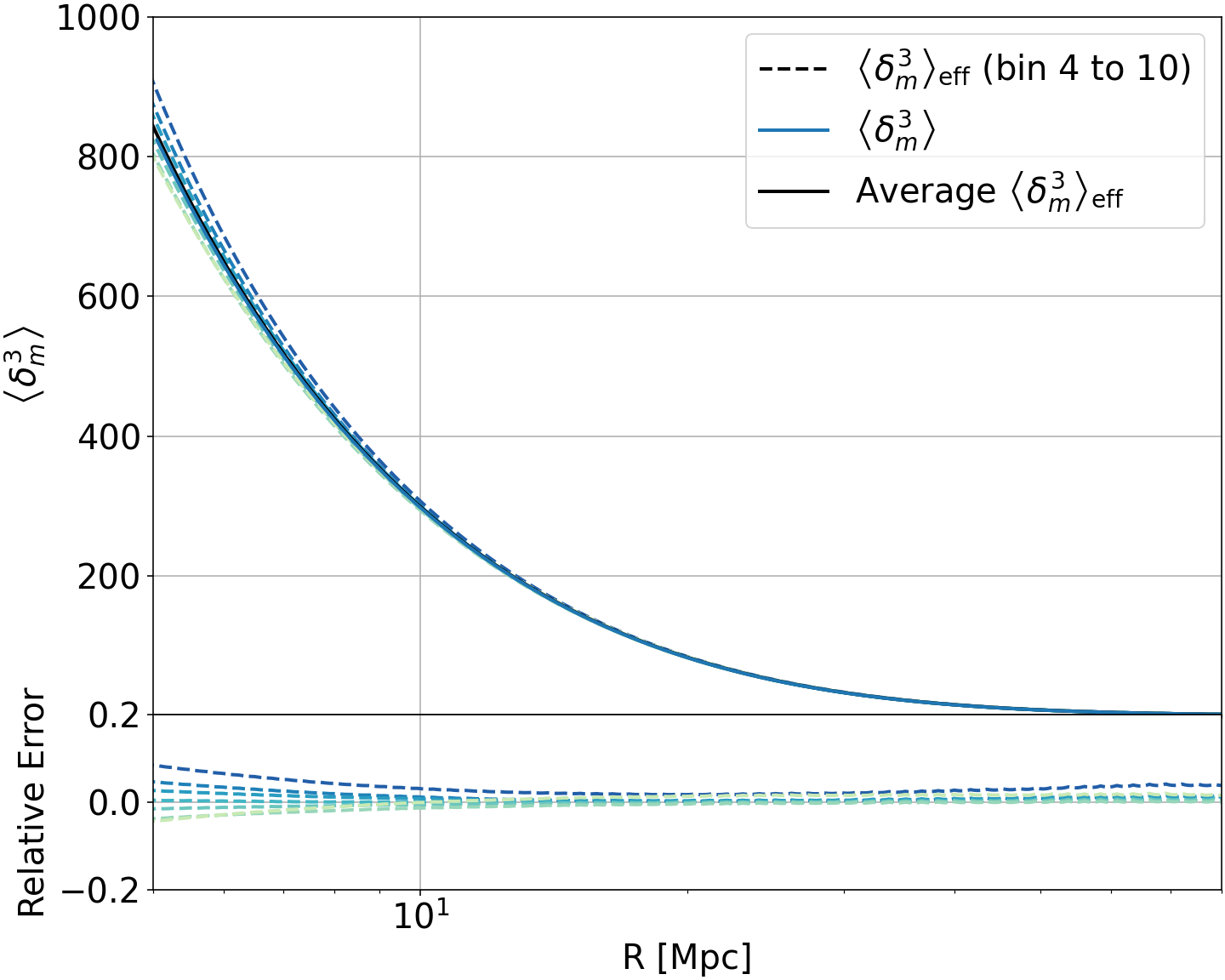}
    \caption{Effective reconstruction of the $\left\langle\delta^3_m\right\rangle$ in real space from bins 3 to 9. Each dashed curve corresponds to a diagonal bin contribution; the solid black curve shows their average, and the solid blue line shows the original $\left\langle\delta^3_m\right\rangle$ used as an input for computing $\left\langle\kappa^3\right\rangle$.}
    \label{fig:deltam3_eff_reconstruction}
\end{figure}

As an illustration of the previous reconstruction, we here reconstruct the matter density skewness smoothed in an infinitely long cylinder of (varying) radius $R = \chi_{\rm eff}\theta$ and at chosen redshift $z_{\rm ref} = 1$. Its theoretical expression from perturbation theory is given in Appendix~\ref{sec:deltam3comp} and its reconstructed expression from the BNT convergence field in bin $a$ is obtained through Eq.~\eqref{eq:cumulant} which now reads
\begin{equation}
    \left\langle\delta_m^3\right\rangle_\text{eff} = \frac{\left\langle\BNT{\kappa}_a^3\right\rangle}{\left(\BNT{\omega}_a(\chi_{\text{eff},a})\right)^3} \times \sqrt{\frac{3}{2 \pi \sigma_{\chi,a}^2}} \times \frac{D_+^4(z_{\rm ref})}{D_+^4(z_{\chi_{\text{eff},a}})}.
\end{equation}
We illustrate this reconstruction at $z_{\rm ref}=1$ in Fig.~\ref{fig:deltam3_eff_reconstruction}.
The reconstruction is there very convincing with little to no dispersion across source bins and constant residuals with respect to the theoretical expectation for scales above $\sim10$ Mpc where we expect perturbation theory to hold.

\section{Nonlinear evolution} 
\label{sec:non_linear}

The previous reconstruction method, be it at the probe or field level, assumed linear evolution, or some variants in the form of simple couplings of linear fields from perturbation theory, as in the case of our reconstruction of the matter density skewness. However, although those expressions and presentation enable accurate and analytical reconstructions, the ability to recover the matter density field from tomographic lensing experiments is geometric in nature, as is the BNT transform. This should in principle enable one to plug in any model for the growth of structure. 

The nonlinear couplings between scale and redshift evolution could however spoil the interest of the method since each BNT source bin technically probes a different scale(s) and time, and since the proposed dynamical rescaling was only a way to output an effective observable defined as the mean of all the rescaled observables in each bin. Nevertheless, one could define a generalized \textit{ad hoc} nonlinear growth factor defined as the ratio of the target observable at the probed scale(s) and time to the same quantity at the target scale(s) and redshift. A simple example can be constructed for the matter power spectrum at the probe level, and in a very similar fashion at the field level. Let us introduce
\begin{equation}
    W_{\text{eff},nl,ab}(\ell)=\int \frac{\dd \chi}{\chi^2} \BNT{\omega}_a(\chi)\BNT{\omega}_b(\chi) \frac{D_{+,nl}^2\left(\chi,\frac{\ell}{\chi}\right)}{D^2_{+,nl}\left(\chi(z_{\rm ref}),\frac{\ell}{\chi}\right)},
\end{equation}
enabling us to define
\begin{equation}
    P_{\text{eff},nl,ab}(k)=\frac{\mathcal{C}^{\BNT{\kappa} \BNT{\kappa}}_{ab}(k \left< \chi \right>_{ab}) }{W_{\text{eff},nl,ab}(k \left< \chi \right>_{ab})} \approx P(k,z_{\rm ref}).
\end{equation}

\begin{figure}[!ht]
    \centering
    \begin{subfigure}{\columnwidth}
        \centering
        \includegraphics[width=0.99\linewidth]{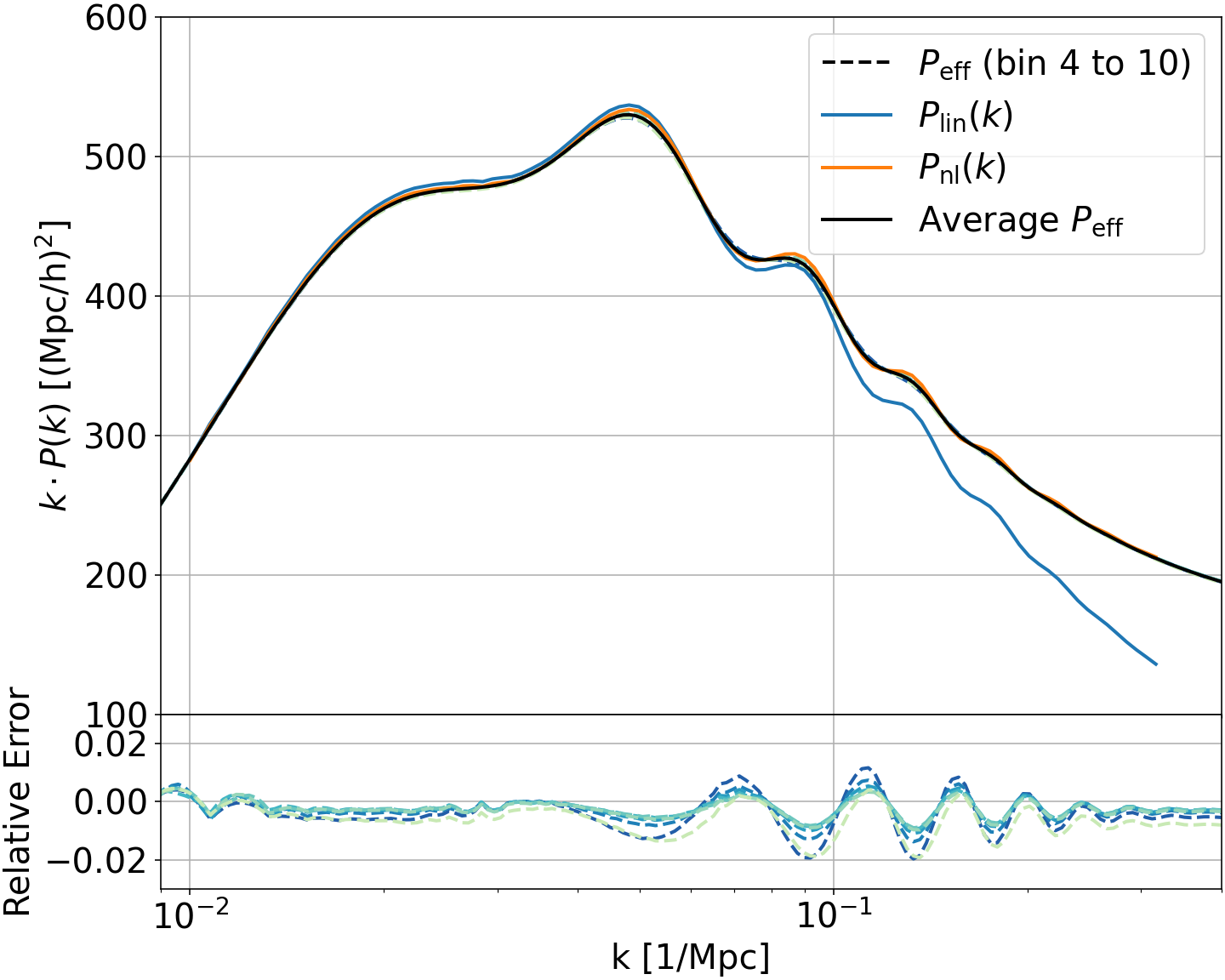}
        \caption{Rescaled nonlinear effective power spectrum $k \cdot P_{\mathrm{eff},nl}(k)$ for each bin and their average.}
        \label{fig:Peffnl}
    \end{subfigure}

    \begin{subfigure}{\columnwidth}
        \centering
        \includegraphics[width=0.99\linewidth]{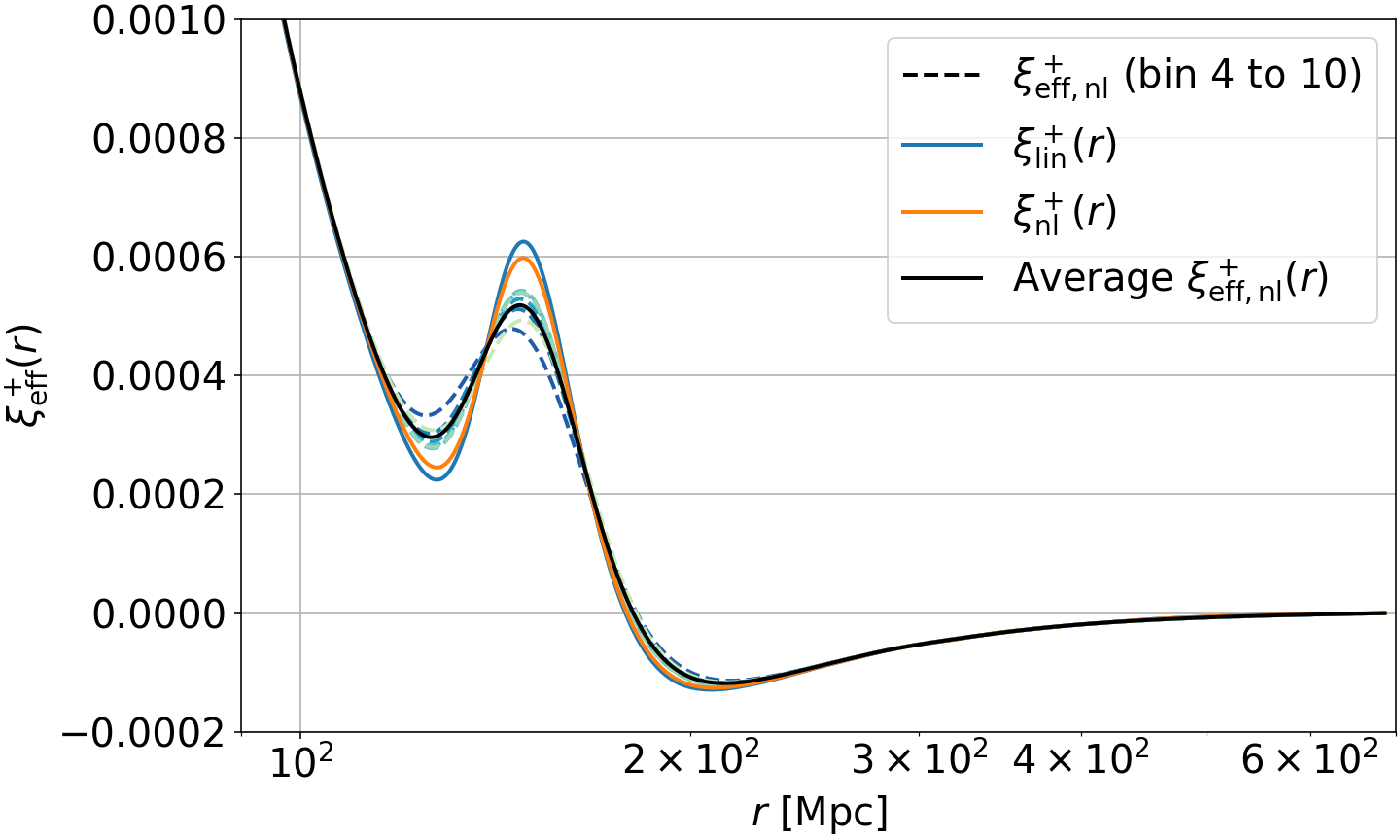}
        \caption{Effective nonlinear two-point correlation function $\xi^+_{\mathrm{eff}}(r)$ reconstructed from $P_{\mathrm{eff},nl}$.}
        \label{fig:xieffnl}
    \end{subfigure}

    \caption{Probe-level nonlinear effective power spectrum and its real-space counterpart in single BNT bins and nonlinear evolution of the rescaling.}
    \label{fig:Peff_xi_nl_combined}
\end{figure}

The obtained effective matter power spectrum at $z_{\rm ref}=1$ is displayed in Fig.~\ref{fig:Peffnl}.

Figure~\ref{fig:Peff_xi_nl_combined} illustrates that using a nonlinear growth enables a precise reconstruction, here with a relative error below $2\%$ for a large range of $k$ values. Moreover, the relative error tends to be larger at the locations of the wiggles while remaining below $1 \%$ at scales away from the wiggle region. Thus, the main driver of the fractional difference is the damping of the baryonic oscillations which is a consequence of the finite width of the BNT-transformed lensing kernels. It is nevertheless hard to find substantial gain from the implementation of nonlinear growth at least in this example apart from reduced dispersion, compared to Fig.~\ref{fig:Peff_xi_combined}, from the effective curve coming from all source bins, which does illustrate a more precise reconstruction.

\section{Conclusion}
\label{sec:conclusion}

This paper presents novel approaches in tomographic cosmic shear analysis designed to efficiently probe the underlying three-dimensional matter density contrast. The analyses conducted here are all based on the BNT transform \cite{Bernardeau:2013rda}, whether applied at the “probe level” in the first half of the paper—specifically to summary statistics such as angular power spectra—or at the field level, as explored in greater depth in the second half. In both cases, we exploit the BNT transform’s ability to generate lensing observables with arbitrarily\footnote{There is no mathematical barrier to reducing the lensing kernel as close as one wants to a Dirac delta function, but this is obviously limited in practice by the finite density of galaxies which trace the lensing fields.} narrow lensing kernels for a given source redshift distribution. Within the framework of the Limber approximation, this enables the localization of contributing density fluctuations in both redshift and scale.

When applied to standard summary statistics, BNT-transformed quantities retain the same information content. However, it was quickly realized that this transformation allows for the accurate separation of different physical scales, thereby facilitating the introduction of efficient scale cuts. A notable consequence of applying the BNT transform to individual bins of the source redshift distribution is the large covariance associated with each element of the newly constructed BNT data vector. This limitation is frequently cited in the literature as a major objection to the use of the BNT transform. Yet, as demonstrated in the first half of this paper, these BNT bins are strongly correlated. The reason lies in the more intricate noise structure of the covariance matrix (used here as a first-order estimate of the likelihood) of our transformed data vector. In particular, elements whose expectation value tends to zero—i.e., those with a vanishing cosmological signal—cannot be naively discarded, as they still possess a nontrivial second moment. They must therefore be included in the analysis, without applying specific scale cuts.

This behavior is quantitatively illustrated in the correlation matrix shown in Fig.~\ref{fig:BNTcorr}. We further emphasize this point in Sec.~\ref{sec:detect}, where the figure of merit for BAO detection using the cosmic shear two-point correlation function increases by a factor of $4$ (see Fig.~\ref{fig:BAOFisher}). These results are obtained while accounting for realistic noise sources and implementing scale cuts as would be done in actual data analysis. More broadly, our findings call into question the $\ell$-cut strategies currently employed in surveys, such as those recently presented in Ref.~\cite{Gu:2024izq}. Our analysis suggests that stringent $\ell$-cut strategies applied to data vector elements with vanishing signal-to-noise ratios are suboptimal.

In the second half of the paper, we delve deeper into the structure of the BNT-transformed field. We show that it is possible to construct data vectors in which similar physical scales are grouped together, unlike in standard projected observables. To achieve this, we introduce explicit analytical rescaling of the lensing observables. For instance, estimates of (correlators of) the matter density field can be appropriately rescaled at the “field level” once BNT transformed. The corresponding geometrical factors for the two-point correlation function are provided in Eq.~\eqref{eq:Peffprobe} and for the n-point cumulants and correlators in Eqs.~\eqref{eq:cumulant} and~\eqref{eq:functional1}. These factors ensure that the expectation value of each element of the BNT-transformed data vector either probes the same type of structures (same scale and time) or tends to zero. This allows the entire signal from all sources to be refocused onto a single key scale, demonstrating that the full range of matter density features can, in principle, be observed from tomographic lensing experiments—regardless of any constraints imposed by the covariance matrix (i.e., irrespective of whether the inverse problem is ill posed or requires additional assumptions such as sparsity \cite{Leonard:2013hia}).

Finally, we note that both the data vector and the covariance matrix can be expressed analytically as functions of the usual non-BNT-transformed vector (mathematically corresponding to a change of basis and a rescaling). This opens the possibility of finding an optimal linear combination of the (rescaled) noisy data vector to reproduce, for example, Fig.~\ref{fig:Peff_xi_combined} (presented here without noise). Such an approach would extend our demonstration and provide a natural consistency test for the estimation of cosmological parameters in galaxy surveys, potentially enabling the direct observation of features such as the BAO peak. We leave this investigation for future work.

\begin{acknowledgments}
This work has received funding from the Centre National d’Etudes Spatiales for travel and hardware. This work was initiated during the workshop YITP-T-24-05 at the Yukawa Institute for Theoretical Physics, we thank the organizers for this particularly useful and interesting workshop.
\end{acknowledgments}

\section{Data Availability}

There are no publicly available research data or software supporting this manuscript. Requests for further information of data should be sent to authors.

\bibliography{biblio}

\appendix

\section{Specifications of our synthetic survey}\label{sec:euclidspec}

Throughout this work, we use a Euclid-like setting as specified by \cite{Euclid:2019clj,Deshpande:2019sdl} such that the galaxy number density is given by
\begin{equation}
    n(z)\propto \left(\frac{z}{z_0}\right)^2 \exp \left[-\left(\frac{z}{z_0}\right)^\frac{3}{2}\right],
    \label{eq:n_z}
\end{equation}
where $z_0=z_{\rm m}/\sqrt{2}$ and $z_{\rm m}=0.9$ is the median redshift of the survey. We also consider an overall galaxy surface density of $30 {\rm \ arcmin^{-2}}$, which here only impacts the shape-noise amplitude. We then build the redshift bins as
\begin{equation}
n_i(z)=\frac{\int_{z_i}^{z_{i+1}} \dd z_p\, n(z) p_{\rm ph}(z_p,z)}{\int_{z_{min}}^{z_{max}} \dd z \int_{z_i}^{z_{i+1}} \dd z_p\, n(z) p_{\rm ph}(z_p,z)},
\label{eq:n_iz}
\end{equation}
where $p_{\rm ph}$ is the probability density that a galaxy at redshift $z$ is measured at redshift $z_p$. We use here a simplified version of the probability density from \cite{Deshpande:2019sdl} for which we have
\begin{equation}\label{eq:photoprob}
p_{\rm ph}(z_p,z)= \frac{\exp\left[-\frac{1}{2} \left(\frac{z-z_p}{\sigma_z (1+z)}\right)^2\right]}
{\sqrt{2 \pi} \sigma_z (1+z)} ,
\end{equation}
where $\sigma_z=0.02$ is the variance of the photometric redshift measurement at redshift $0$.
We chose to work with equipopulated bins in order not to statistically favor any epoch and consider the usual Euclid-like setup with ten tomographic redshift bins as stated in Ref.~\cite{Euclid:2019clj}. Finally, we consider a shape noise with a flat power spectrum and a realistic standard deviation on individual galaxies of $\sigma_s = 0.3$.

\section{Tomographic Precision requirement for BAO measurement from cosmic shear}\label{sec:tomoprec}

Tomography is a necessary feature for measuring BAO using cosmic shear. The tomographic precision of an experiment therefore determines its performance in accessing BAO features. In this section, we quantify and illustrate this requirement on tomographic precision. Thus, we determine the required precision in $k$ that enables the detection of BAO wiggles in cosmic shear, and we relate it to tomographic precision.

\begin{figure}
    \centering
    \includegraphics[width=0.99\linewidth]{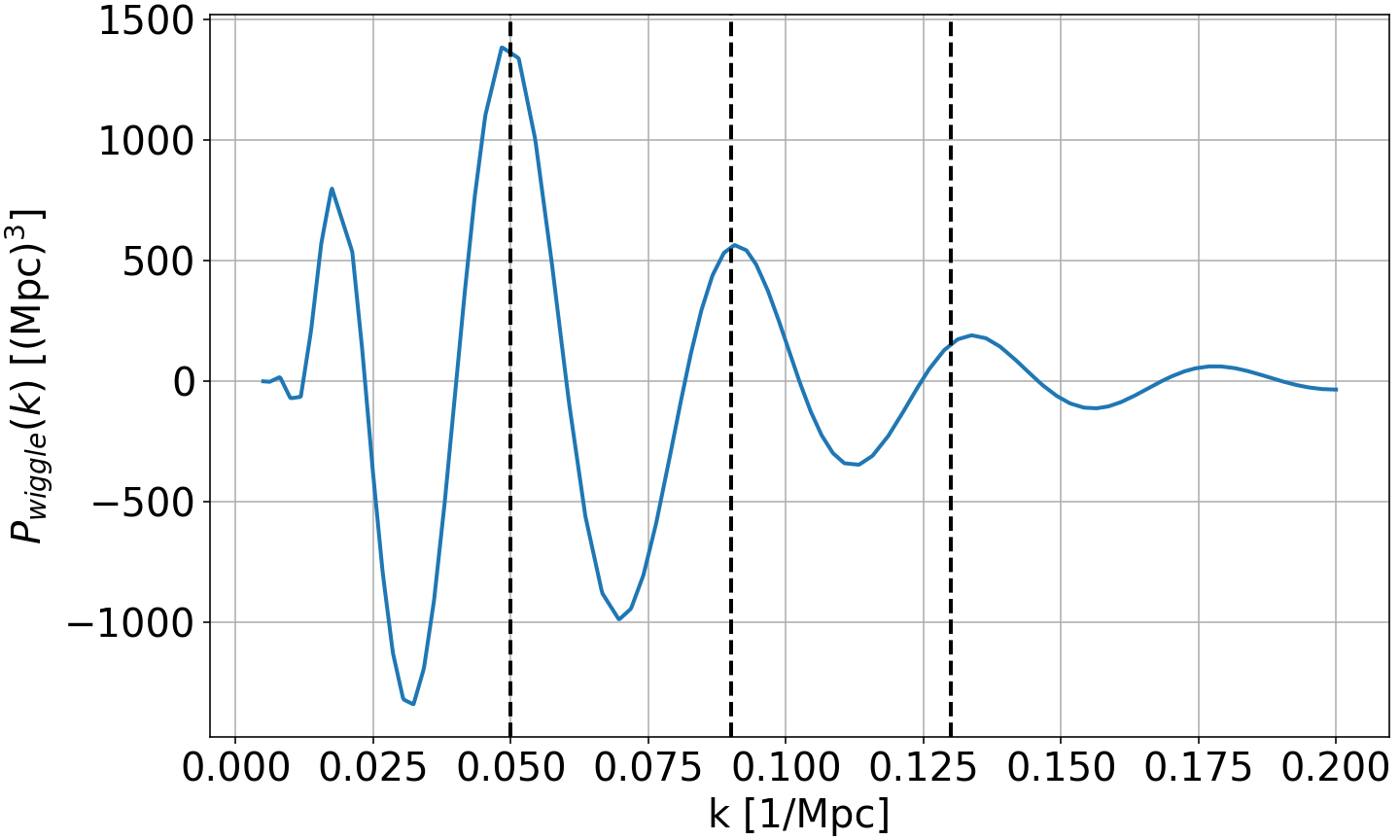}
    \caption{Wiggles of the matter power spectrum obtained from CLASS \cite{Blas:2011rf} using the linear matter power spectrum and its numerical no-wiggle counterpart. Vertical dotted lines are used to show the position $k_{\rm ref}=0.05 {\rm \ Mpc}^{-1}$ and the pseudoperiod $\Delta k_{\rm BAO} \approx 0.042 {\rm \ Mpc}^{-1}$ of the oscillations.}
    \label{fig:Pwiggle}
\end{figure}

First we plot the wiggle part of the matter power spectrum in Fig.~\ref{fig:Pwiggle} as defined in Sec.~\ref{sec:BAO} and denoted as $P_{\rm w}$ and watch for position and pseudowavelength of the wiggles. Thus, we define $k_{\rm ref}=0.05 {\rm \ Mpc}^{-1}$ as the position of the wiggles and their period in wave number would then be $\Delta k_{\rm BAO} = 2 \pi/r_{\rm drag} \approx 0.042 {\rm \ Mpc}^{-1}$. In Ref.~\cite{Bernardeau:2020jtc}, a tomographic precision criterion is given by expressing the $k$ resolution of a reconstructed $P_{\rm eff}$ or $C(\ell)$ from a given kernel as
\begin{equation}
    \Delta k_{\rm eff} = k_{\rm eff} \frac{\Delta \chi_{\rm eff}}{\chi_{\rm eff}},
\end{equation}
where $k_{\rm eff}=l/\chi_{\rm eff}$ is the chosen scale of work, $\chi_{\rm eff}$ is the effective comoving distance in the lensing kernel and $\Delta \chi_{\rm eff}$ is its thickness. With our setup and notation for the BNT-transformed $a$th bin, $\Delta \chi_{\rm eff}=\sigma_{\chi,a}$. Thus, the criterion for being able to observe BAO wiggles in the reconstructed $P_{\rm eff}$ is to have $\Delta \chi_{\rm eff}$ small enough compared to the half-period of BAO wiggles in wave number $\Delta k_{\rm BAO}$. This requirement is mathematically due to a destructive interference phenomenon that comes from the mixing of scale when performing Limber integration over a large lensing kernel.

\begin{figure}
    \centering
    \includegraphics[width=0.99\linewidth]{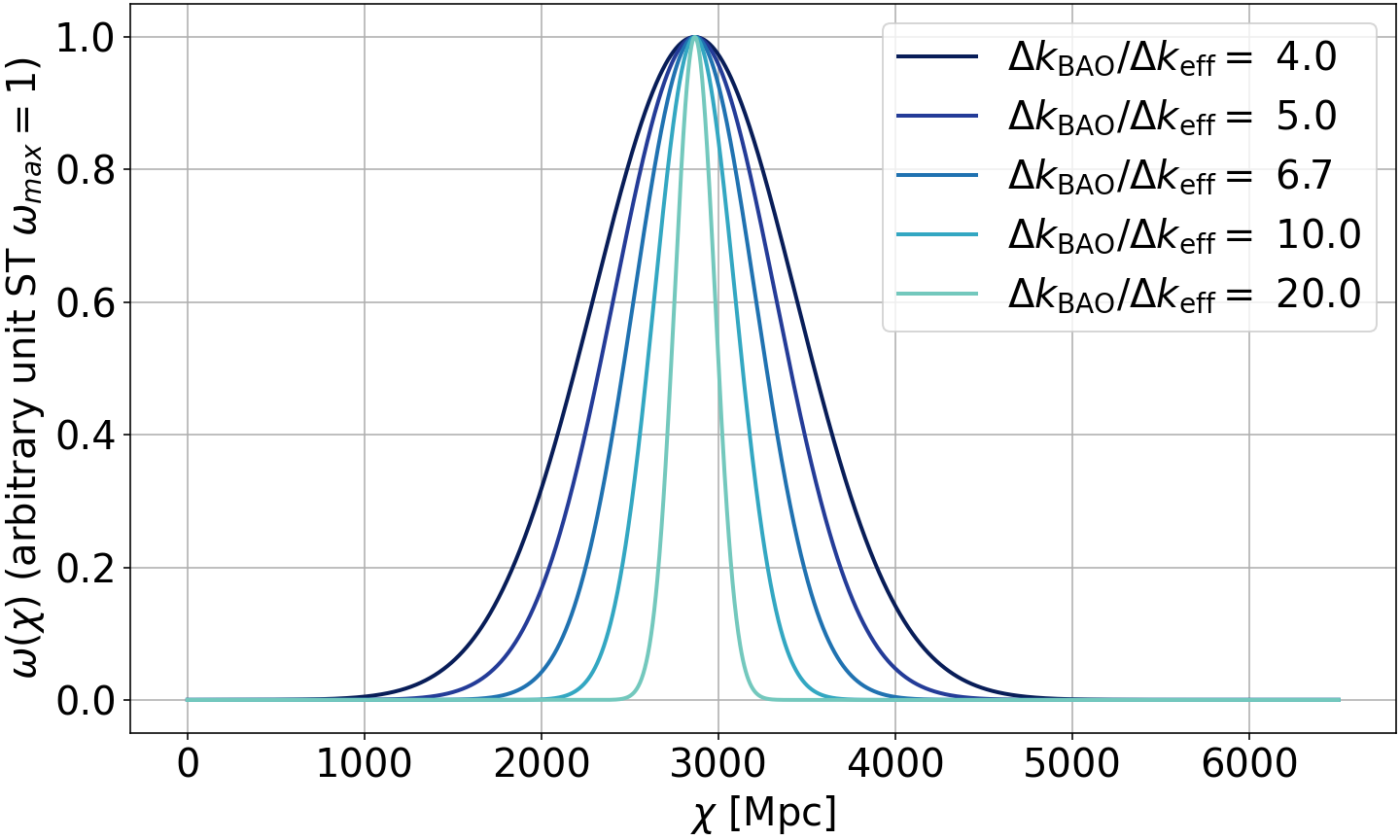}
    \caption{Gaussian BNT-like weak lensing kernels for different thickness.}
    \label{fig:limitkernel}
\end{figure}

We now want to test for this criterion in order to give a tomographic precision requirement. To do so we will compute a reconstructed matter power spectrum $P_{\rm eff}$ for various largeness of lensing kernels which we show in Fig.~\ref{fig:limitkernel}. Those are defined as normal distributions in the space of comoving distance which is a justified procedure given Sec.~\ref{ssec:saddlepoint}. All have mean $\chi_{{\rm eff},0.8}=\chi(z=0.8)$ (redshift $0.8$) with various thicknesses that give various $k$ resolutions at $k_{\rm eff}=k_{\rm ref}$ from $\Delta k_{\rm eff}= 0.01 {\rm \ Mpc}^{-1}\approx \Delta k_{\rm BAO}/4$ to $\Delta k_{\rm eff}= 0.001 {\rm \ Mpc}^{-1}\approx \Delta k_{\rm BAO}/20$, the latter being of the same order of magnitude as what we have in our Euclid-like setup, using ten bins (see Fig.~\ref{fig:winulling}).

\begin{figure}
    \centering
    \includegraphics[width=0.99\linewidth]{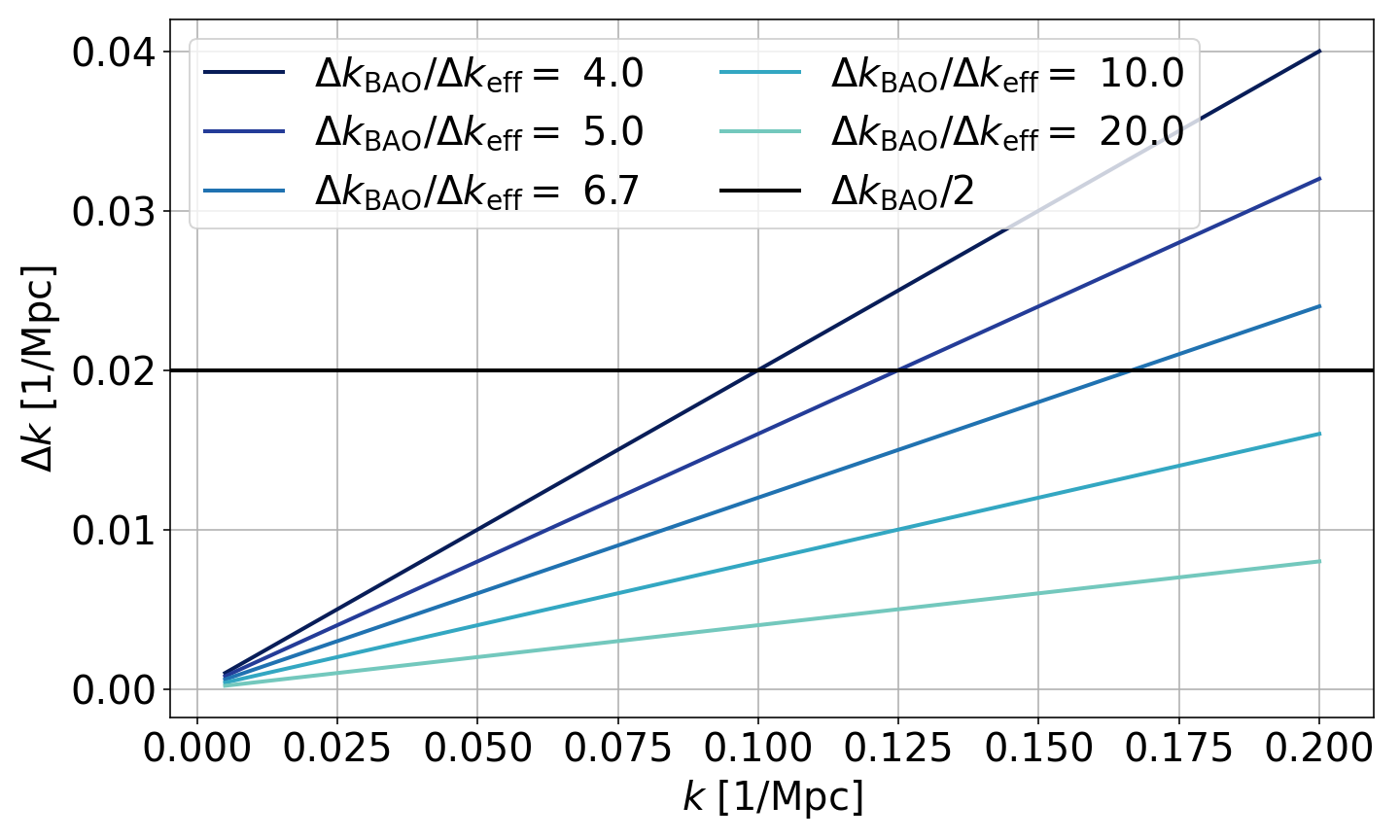}
    \caption{Evolution of the $k$ resolution for each test bin. The horizontal black line shows the resolution under which wiggles are detected.}
    \label{fig:deltak}
\end{figure}

We show in Fig.~\ref{fig:deltak} the evolution of the $k$ resolution for each lensing kernel used in this test. We see here that for our two lest thick kernels, the $k$ resolution is under $\Delta k_{\rm BAO}/2$ for all scales where wiggles are visible in the matter power spectrum. This is not the case for the three others.

\begin{figure}
    \centering
    \includegraphics[width=0.99\linewidth]{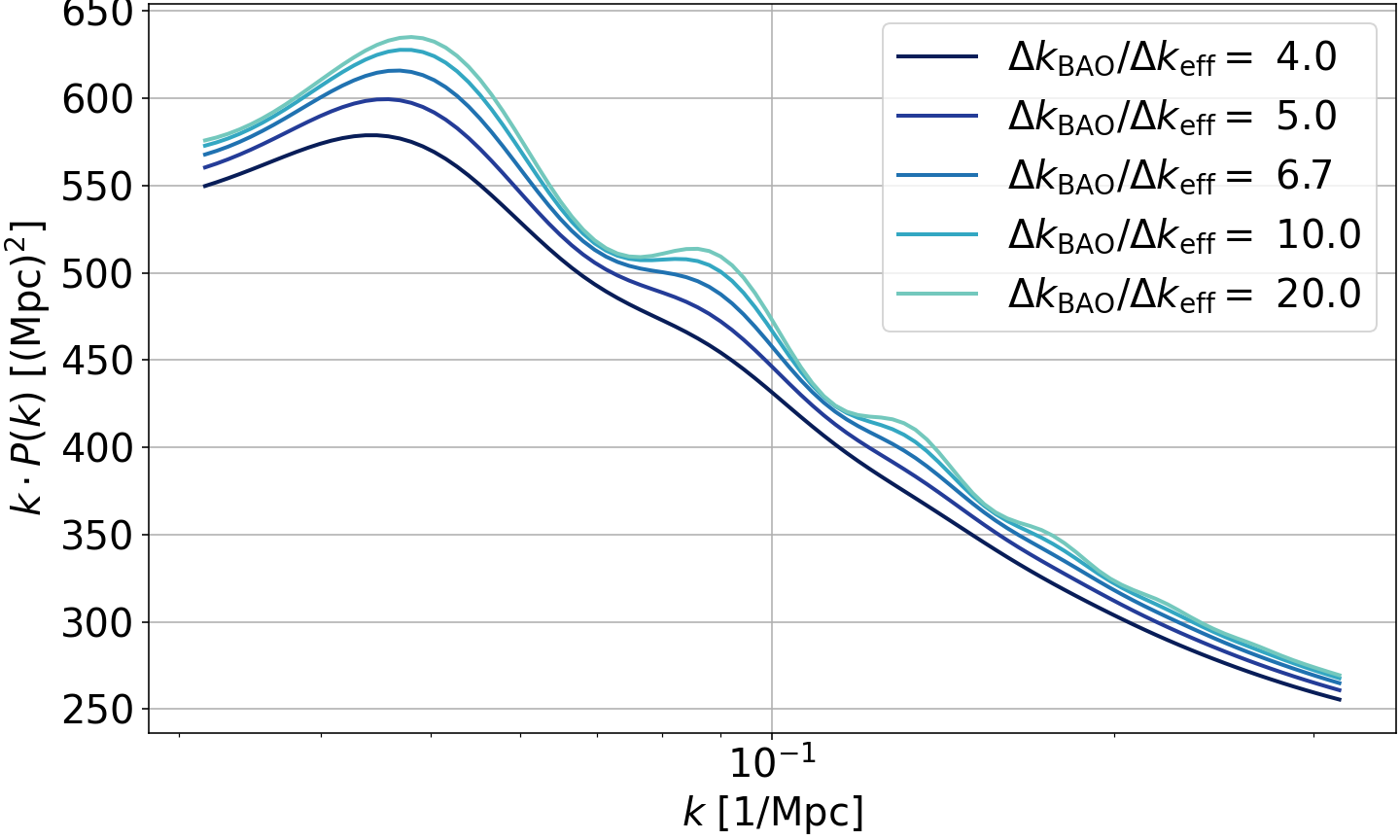}
    \caption{Effective matter power spectrum for each test bin. One can see that the thicker the kernel is, the more wiggles are damped.}
    \label{fig:limitPk}
\end{figure}

We finally show in Fig.~\ref{fig:limitPk} the effective matter power spectrum obtained by performing our reconstruction scheme as in Eq.~\eqref{eq:Peffprobe} at $z_{\rm ref}=0.8$. In this plot, one can very clearly see the damping of BAO wiggles when the thickness lensing kernel increases. There is most likely no way to perform a scale mixing reduction with lensing kernel significantly less thick than what the BNT transform enables. Thus, a suitable criterion on the tomographic precision for BAO detection in the cosmic shear would be that the BNT-transformed lensing kernels should be small enough such that the $k$ resolution in every bins is significantly smaller than the half-period of BAO wiggles for $0.01 {\rm \ Mpc}^{-1}< k < 0.2 {\rm \ Mpc}^{-1}$. Given the evolution of the $k$ resolution with $k$, this criterion becomes $\Delta k_{\rm eff} < \Delta k_{\rm BAO}/20$ at $k_{\rm ref}$ which translates to around ten redshift bins in our Euclid-like setup. This means that this criterion will be satisfied by Stage IV large-scale surveys such as Euclid's last data release \cite{Euclid:2024yrr}.

\section{Limitations from catastrophic photometric redshift errors}\label{sec:photo_error}

One of the potential blind spots of the methods developed in this paper is catastrophic photometric redshift errors which could bias the tomographic bins in a potentially dangerous fashion for cosmological interpretation. In the specific case of BAO measurements, galaxy clustering experiments possess a similar issue \cite{Euclid:2025dlg}. In our case, the impact of redshift precision is of two kinds: Firstly, redshift resolution directly impacts the number of tomographic source bins that can be built which translates in our ability to probe the redshift evolution of observables through the BNT-transformed kernel's width. On the other hand, catastrophic errors, which hence go beyond the simple Gaussian-like widening of the bins, decrease the precision of the effective redshift in each of the source bins and, thus, affect the heart of the method we developed based on the BNT-transformed kernels for weak lensing. Quantitatively, outliers may be defined as galaxies whose true redshifts lie more than three bin widths (standard deviation of source redshifts within the bin) away from the bin mean. Assuming an outlier fraction of $10\%$, roughly Euclid's expectations, a shift of 0.05 in the mean redshift would require the outlier population to be located $0.5$ away from the bin center; see Ref.~\cite{Mill:2025jlc} and our quick derivation below. By comparison, in a Euclid-like analysis with ten tomographic bins, the typical bin width is approximately $0.15$. This means that as long as the proportion of outliers in a typical bin is less than $10\%$, the variation of the effective redshift in the bin of the BNT-transformed kernel is negligible. Thus, in a Euclid-like survey, as long as the redshift calibration meets the requirement of the experiment, the only limitation should be on the number of bins, i.e., the first kind of impact we discussed. In that case, we performed a quantitative investigation of the impact of the width of kernels in the detection of scale-dependent features such as the BAO in Appendix~\ref{sec:tomoprec}. In general, a deeper investigation of systematic effects on our proposed method is left for future work. We quickly derive how the mean redshift in a source bin is affected by a given proportion of outliers, ${\rm f_{outlier}}$:

\begin{equation}
\begin{split}
\left\langle z \right\rangle_{\rm bin} &= (1 - {\rm f_{\, outlier}}) \cdot \left\langle z \right\rangle_{\rm fiducial} + {\rm f_{\, outlier}} \cdot \left\langle z \right\rangle_{\rm outlier} \\ &= \Delta_{\langle z \rangle} + \left\langle z \right\rangle_{\rm fiducial},
\end{split}
\end{equation}
which leads to
\begin{equation}
    \left\langle z \right\rangle_{\rm outlier} = \left\langle z \right\rangle_{\rm fiducial} + {\rm f_{\,outlier}}^{-1}\cdot \Delta_{\langle z \rangle}.
\end{equation}

As a consequence, to shift the mean source redshift in a given bin by $\Delta_{\langle z \rangle} = 0.05$, and with a fraction of outliers ${\rm f_{outlier}} = 0.1$, the mean redshift of the outliers, and assuming the worst case where they are all biased in the same direction, is 0.5 away from the fiducial bin value, almost 3 times further than the Euclid estimate.

\section{Matter density skewness} \label{sec:deltam3comp}

Following \cite{Bernardeau:2001qr} and at leading order in Eulerian perturbation theory, the matter density skewness $\langle\delta_{m}^3\rangle$ smoothed in an infinitely long cylinder (2D top-hat window function) of radius $R$ is given by
\begin{align}\label{eq:deltam3}
    \left\langle\delta_{m}^3\right\rangle=\sigma^4(R,z)\left(3 \nu_2 + \frac{3}{2} \frac{\partial \log  \sigma^2(R,z)}{\partial \log R}\right),
\end{align}
where $\nu_2\sim1.4$ is the spherically averaged 2D $F_2$ perturbation theory kernel,
\begin{equation}
    \sigma^2(R,z) = \int \frac{\dd \mathbf{k}_\perp}{(2 \pi)^2} (WTH(k_\perp R))^2 P(k_\perp,z),
\end{equation}
$WTH(x)=2 J_1(x)/x$, and $J_1$ is the first Bessel function of the first kind.
\end{document}